\documentclass[preprint]{elsarticle} 

\makeatletter
	\def\ps@pprintTitle{%
 	\let\@oddhead\@empty
	\let\@evenhead\@empty
	\def\@oddfoot{\centerline{\thepage}}%
	\let\@evenfoot\@oddfoot}
\makeatother

\usepackage{etoolbox}
\patchcmd{\MaketitleBox}{\footnotesize\itshape\elsaddress\par\vskip36pt}{\footnotesize\itshape\elsaddress\par\parbox[b][36pt]{\linewidth}{\vfill\hfill\textnormal{\today}\hfill\null\vfill}}{}{}%
\patchcmd{\pprintMaketitle}{\footnotesize\itshape\elsaddress\par\vskip36pt}{\footnotesize\itshape\elsaddress\par\parbox[b][36pt]{\linewidth}{\vfill\hfill\textnormal{\today}\hfill\null\vfill}}{}{}%

\usepackage[colorlinks=true,%
            breaklinks=true,%
            linkcolor=blue,
            urlcolor=blue,%
            citecolor=blue,%
            pdftitle={Title of paper}, 
            pdfkeywords={Keywords},	
            pdfauthor={Authors},		
            bookmarksopen=false,
            pdfpagemode=UseNone]{hyperref}

\usepackage[margin=0.95in]{geometry}
\usepackage[T1]{fontenc}
\usepackage[english]{babel}
\usepackage{cancel}
\usepackage{siunitx}
\usepackage{icomma}
\usepackage[latin1]{inputenc} 
\usepackage{subfiles}
\usepackage{tabularray}
\biboptions{numbers,sort&compress}
\usepackage{subcaption}
\usepackage[justification=centering]{subcaption}

\usepackage{amsmath}
\usepackage{pifont}

\usepackage{mathtools}
\usepackage{siunitx}
\usepackage{bbm}
\usepackage{amsmath}
\usepackage{amsfonts}
\usepackage{amsthm}
\usepackage{amssymb}
\usepackage{array}
\usepackage{algorithm} 
\usepackage{algorithmicx}
\usepackage{algpseudocode}
\usepackage{subcaption}
\usepackage{siunitx}
\usepackage{amsmath}
\usepackage{mathrsfs}
\usepackage{xpatch}

\usepackage{color}
\usepackage{url}
\usepackage{cleveref}
\usepackage{graphicx}
\usepackage{breakcites}
\usepackage{soul} 
\usepackage{enumitem}
\usepackage[title,titletoc,toc]{appendix}

\newtheoremstyle{mytheoremstyle}{5pt}{5pt}{\itshape}{}{\bfseries}{.}{.5em}{} 
\theoremstyle{mytheoremstyle}

\newtheoremstyle{myremarkstyle}{3pt}{3pt}{\itshape}{}{\bfseries}{.}{.5em}{} 
\theoremstyle{myremarkstyle}

\usepackage{pifont}

{\noindent {\textbf{Proof}:} }%
{\hfill $\Box$ \\ }

\newcommand{\bit}{\begin{itemize}}
\newcommand{\eit}{\end{itemize}}
\newcommand{\ben}{\begin{enumerate}}
\newcommand{\een}{\end{enumerate}}

\newcommand{\Hquad}{\hspace{0.5em}}

\begin{document}
\begin{frontmatter}
    \title{Conservative projection-based data-driven model order reduction of a fluid-kinetic spectral solver}
    \author[1,2]{Opal Issan\corref{cor1}}\ead{oissan@ucsd.edu}
    \author[2]{Oleksandr Koshkarov}
    \author[3]{Federico D. Halpern}
    \author[2]{Gian Luca Delzanno}
    \author[1]{Boris Kramer}
    
    \cortext[cor1]{Corresponding author}
    \address[1]{Department of Mechanical and Aerospace Engineering, University of California San Diego, La Jolla, CA, USA}
    \address[2]{T-5 Applied Mathematics and Plasma Physics Group, Los Alamos National Laboratory, Los Alamos, NM, USA}
    \address[3]{General Atomics, P.O. Box 85608, San Diego, CA, USA}

    \begin{abstract}
    Kinetic simulations are computationally intensive due to six-dimensional phase space discretization. Many kinetic spectral solvers use the asymmetrically weighted Hermite expansion due to its conservation and fluid-kinetic coupling properties, i.e., the lower-order Hermite moments capture and describe the macroscopic fluid dynamics and higher-order Hermite moments describe the microscopic kinetic dynamics. We leverage this structure by developing a parametric data-driven reduced-order model based on the proper orthogonal decomposition, which projects the higher-order kinetic moments while retaining the fluid moments intact. We demonstrate analytically and numerically that the method ensures local and global mass, momentum, and energy conservation. The numerical results show that the proposed method effectively replicates the high-dimensional spectral simulations at a fraction of the computational cost and memory, as validated on the weak Landau damping and two-stream instability benchmark problems.
    \end{abstract}	
\end{frontmatter}

\section{Introduction}\label{sec:introduction}
A kinetic description is essential to accurately capture the behavior of many collisionless plasma phenomena, including magnetic reconnection, sheaths, shocks, and wave-particle interactions. The kinetic equations evolve the particle distribution function in a six-dimensional 3D3V phase space. Solving these equations via an Eulerian discretization on a grid approach is computationally demanding. The memory costs scale as $\mathcal{O}(N^{6})$, where the dimension $N$ represents the number of degrees of freedom~(DOFs) in each spatial or velocity dimension~(although, in practice, the number of DOFs might differ in each dimension). Such unfavorable scaling is commonly known as the \textit{curse of dimensionality}. Moreover, the often enormous scale separation between microscopic and macroscopic processes renders kinetic simulations computationally intractable at large scales, especially for many-query parametric studies, including uncertainty quantification, sensitivity analysis, inverse problems, and design optimization.

A promising avenue to reduce these computational costs is to develop kinetic reduced-order models~(ROMs), most commonly done via modal decomposition techniques. Within this class of methods, there are two main reduction approaches for the kinetic equations: proper orthogonal decomposition~(POD) and dynamic low rank~(DLR). 
The latter approximates the high-dimensional particle distribution function by a low-dimensional modal decomposition that results in storage requirements that scale as~$\mathcal{O}(N_{r} N^3)$ and computational efforts that scale as~$\mathcal{O}(N_{r}^2 N^3)$, where $N_{r}$ is the rank of the approximation~\cite{cassini_2022_dlr_3d3v, othmar_2007_dlr}. DLR evolves not only the reduced basis coefficients but also the reduced basis over time, such that this method does not rely on an offline phase, in contrast to POD. 
Here, the offline phase refers to expensive pre-computations (e.g., basis generation), while the online phase leverages this pre-computed information for fast and efficient evaluations.
A comprehensive review of DLR for the kinetic equations is provided in~\citet{einkemmer_2024_review}, including recent advancements that enabled 3D3V simulations on a single workstation~\cite{cassini_2022_dlr_3d3v}. 
Alternatively, POD relies on \textit{a priori} training simulation data to construct a global parametric basis using the singular value decomposition, i.e., the offline phase~\cite{lumley_1967_structure, berkooz_1993_pod}. The system dynamics are then projected onto the POD basis and evolved in the reduced linear subspace.
POD was independently rediscovered several times by~\citet{pearson_1901_pca, hotelling_1933_pca, loeve_1955_pca, karhunen_1946_pca}, and is also referred to as the Karhunen-Lo{\`e}ve expansion or Principal Component Analysis.
This method evolves the global basis coefficients, trading higher offline computational costs for a more efficient reduced model during the online phase. Therefore, POD is very useful for parametric studies, where similar problems are solved sequentially. 
The POD (without hyper-reduction) online storage requirements scale as $\mathcal{O}(N_{r})$, while the computational efforts scale as $\mathcal{O}(N_{r}^{2} N^3)$ due to the Vlasov nonlinearity, where $N_{r}$ represents the rank of the reduced POD basis. The memory footprint is significantly lower for POD in comparison to DLR. Additionally, although the POD and DLR computational efforts scale similarly,~\citet{koellermeier_2024_micro_macro_shallow_water} show that for the shallow water equations, POD provides a speed-up factor of 50, while DLR achieves a speed-up factor of around 8 at the same error level. This is anticipated because POD relies on a fixed basis, whereas DLR requires solving additional equations to update the reduced basis.
In several works, coherent structures and limit cycle identification in plasma simulations have been explored by analyzing the POD basis~\cite{sasaki_2019_turbulence, kaptanoglu_2021_rom_mhd, gahr_2024_opinf}. POD has also been applied to particle-in-cell kinetic simulations in~\citet{julio_2019_pic} and its symplectic formulation in~\citet{tyranowski_2022_pic}. POD is also employed in the design of preconditioners to enhance the convergence rate of iterative solvers in high-dimensional simulations~\cite{peng_2024_preconditioner}. Additionally,~\citet{tsai_2023_rom_vp} use POD to reduce a conservative Eulerian finite difference discretization of the 1D1V collisionless Vlasov-Poisson equations. However, the resulting ROM fails to preserve mass, momentum, and energy conservation laws, which are essential for accurately predicting macroscopic dynamics. This paper addresses this challenge.

We propose a novel strategy for constructing data-driven conservative parametric kinetic ROMs. The method is based on the asymmetrically weighted Hermite spectral expansion in velocity~\cite{grant_1967_hermite, schumer_1998_sw_aw, camporeale_2016_sps}. The so-called \textit{fluid-kinetic} coupling in this formulation arises from the lower-order Hermite expansion coefficients representing macroscopic fluid dynamics, while the higher-order coefficients account for microscopic kinetic effects. Leveraging this fluid-kinetic coupling property, we only build a ROM for the higher-order moment equations, as most memory and computational resources are spent on resolving the kinetic moments. As a result, the fluid moments remain intact, ensuring local and global conservation of mass, momentum, and energy.
%
%
The concept of selectively reducing only the higher-order moments of the dynamics was introduced in~\citet{peng_2021_jcp_micro_macro_radiation} using DLR for the radiation transport equations and extended to the kinetic equations in~\citet{coughlin_2024_jcp_micro_macro} using DLR.
Furthermore,~\citet{koellermeier_2024_micro_macro_shallow_water} applied POD and DLR reduction to higher-order moments in the shallow water equations, demonstrating both local and global mass conservation.
We build on this idea by employing POD for the higher-order moments of the kinetic plasma equations, an approach that, to the best of our knowledge, has not been explored before.
We test the ROM on 1D1V Vlasov-Poisson benchmark problems, including weak Landau damping and two-stream instability, demonstrating a significant reduction in the number of DOFs. 
We anticipate the ROM will be particularly useful in reducing 3D3V simulations due to POD's favorable memory and computational scaling.

The remainder of the paper is organized as follows. Section~\ref{sec:fom} describes the Vlasov-Poisson equations and their discretization via an asymmetrically weighted Hermite basis in velocity and second-order central finite differencing in space. Section~\ref{sec:rom} presents the data-driven ROMs approach based on POD and its conservation properties. Section~\ref{sec:numerical-results} presents the numerical results, and we conclude the paper in section~\ref{sec:conclusions}.

\section{Vlasov-Poisson Fluid-Kinetic Spectral Solver}\label{sec:fom}
We outline the 1D1V collisionless Vlasov-Poisson equations in section~\ref{sec:vlasov_poisson}. The asymmetrically weighted Hermite expansion in velocity is detailed in section~\ref{sec:velocity_hermite}, while section~\ref{sec:artificial_collisions} discusses the artificial collision operator introduced in~\citet{camporeale_2016_sps}. Lastly, section~\ref{sec:vector_form} describes the spatial discretization via second-order central finite differences and reformulates the semi-discrete equation in vector form.

\subsection{Vlasov-Poisson Equations}\label{sec:vlasov_poisson}
We study the 1D1V Vlasov-Poisson equations describing the behavior of electrostatic (unmagnetized) collisionless non-relativistic plasma. The normalized Vlasov-Poisson equations evolve the particle distribution function $f_{s}(x, v, t)$ of species $s$ (e.g., electrons and ions) and electric field $E(x, t)$, such that 
\begin{align}
    \left(\frac{\partial}{\partial t} + v \frac{\partial}{\partial x}+ \frac{q_{s}}{m_{s}}E(x, t)\frac{\partial}{\partial v}\right) f_{s}(x, v, t)  &= 0, \label{vlasov-continuum}\\
    \frac{\partial E(x, t)}{\partial x} &= \sum_{s} q_{s} \int_{\mathbb{R}} f_{s}(x,v, t) \mathrm{d}v, \label{poisson-continuum}
\end{align}
where $q_{s}/m_{s}$ is the charge/mass of species $s$. We consider an unbounded velocity coordinate $v \in \mathbb{R}$, a periodic spatial coordinate $x \in [0, \ell]$, where $\ell$ is the length of the spatial domain, and time $t \geq 0$. We impose a unique solution by enforcing that $\int_{0}^{\ell} E(x, t) = 0$. All quantities in the Vlasov-Poisson equations~\eqref{vlasov-continuum}--\eqref{poisson-continuum} are normalized as follows:
\begin{equation*}
   q_{s} \coloneqq \frac{q_{s}^{d}}{e}, \qquad m_{s} \coloneqq \frac{m_{s}^{d}}{m_{e}}, \qquad t \coloneqq t^{d} \omega_{pe}, \qquad x \coloneqq \frac{x^{d}}{\lambda_{D}}, \qquad v \coloneqq \frac{v^{d}}{v_{te}}, \qquad f_{s} \coloneqq f_{s}^{d}\frac{v_{te}}{n_{e}},  \qquad E \coloneqq E^{d} \frac{e \lambda_{D}}{T_{e}}, 
\end{equation*}
where the superscript `$d$' indicates the dimensional quantities in Gaussian-cgs units, $e$ is the positive elementary charge, $\omega_{pe} \coloneqq \sqrt{4 \pi e^2 n_{e}/m_{e}}$ is the electron plasma frequency, $m_{e}$ is the electron mass, $n_{e}$ is the reference electron density, $v_{te} \coloneqq \sqrt{T_{e}/m_{e}}$ is the electron thermal velocity, $T_{e}$ is a reference electron temperature, $\lambda_{D} \coloneqq v_{te}/\omega_{pe} = \sqrt{T_{e}/4\pi e^{2}n_{e}}$ is the electron Debye length.

\subsection{Asymmetrically Weighted Hermite Spectral Discretization in Velocity}\label{sec:velocity_hermite}
We approximate the particle distribution function $f_{s}(x, v, t)$ via a truncated asymmetrically weighted (AW) Hermite spectral expansion, i.e.
\begin{equation}
    f_{s}(x, v,t) \approx \sum_{n=0}^{N_{v}-1} C_{s, n}(x, t) H_{n}(v; \alpha_{s}, u_{s}), \label{expansion-seperation-of-variables}
\end{equation}
where the two Hermite parameters $u_{s} \in \mathbb{R}$ and $\alpha_{s} \in \mathbb{R}_{>0}$ are tunable parameters that can significantly improve the Hermite spectral convergence~\cite{pagliantini_2023_adaptive}. More specifically, $u_{s}$ is the velocity shifting parameter and $\alpha_{s}$ is the velocity scaling parameter. 
The AW Hermite basis function $H_{n}(v; \alpha_{s}, u_{s})$ is defined in Eq.~\eqref{hermite-basis-function-definition}, see~\ref{sec:Appendix-A} for more details. The Hermite orthogonal basis is well-suited for Maxwellian-like distributions since it is constructed using Hermite polynomials weighted by a Maxwellian distribution.
Due to the basis resemblance to a Maxwellian distribution (thermodynamic equilibrium), $u_{s}$ mimics the characteristic mean flow, and $\alpha_{s}$ mimics the thermal velocity.
By inserting the spectral approximation in Eq.~\eqref{expansion-seperation-of-variables} in the Vlasov equation~\eqref{vlasov-continuum}, multiplying against $H^{m}(v; \alpha_{s}, u_{s})$ defined in Eq.~\eqref{hermite-counter-basis-function-definition}, integrating with respect to $v$, and exploiting the orthogonality and recurrence relation of the AW Hermite basis functions in Eqns.~\eqref{orthogonal-property}--\eqref{identity-1}, we derive a system of partial differential equations for the AW Hermite expansion coefficients:
\begin{equation}\label{f-vlasov-pde}
    \frac{\partial C_{s, n}(x, t)}{\partial t} = -\frac{\partial}{\partial x}\left[\alpha_{s} \sigma_{n} C_{s, n-1}(x, t) + \alpha_{s} \sigma_{n+1} C_{s, n+1}(x, t) + u_{s}  C_{s, n}(x, t) \right] +\frac{2q_{s} \sigma_{n}}{m_{s}\alpha_{s}} E(x, t) C_{s, n-1}(x, t), 
\end{equation}
where $\sigma_{n} \coloneqq \sqrt{n/2}$. We adopt the convention that $C_{s, n<0}= 0$ and impose the \textit{closure by truncation}, i.e. $C_{s, N_{v}}= 0$. Similarly, inserting the spectral approximation in Eq.~\eqref{expansion-seperation-of-variables} in the Poisson equation~\eqref{poisson-continuum} results in 
\begin{equation}\label{f-poisson-pde}
    \frac{\partial E(x, t)}{\partial x} = \sum_{s} q_{s} \alpha_{s} C_{s, 0}(x, t).
\end{equation}
The first three fluid moments, i.e., mass, momentum, and energy, are described by combinations of the first three Hermite expansion coefficients, which we show in more detail in section~\ref{sec:conservation_properties}. 

\subsection{Artificial Collisions}\label{sec:artificial_collisions}
We add an artificial collisional operator to solve the problem of \textit{filamentation}, the cascade of very small-scale structures in the distribution function. 
These highly localized and rapidly varying small-scale structures in velocity space can surpass the resolution capabilities of any method, leading to potential numerical instabilities and recurrence phenomena (artificial temporal periodicity)~\cite{issan_2024_collisions}. 
We employ the collisional operator introduced in~\citet{camporeale_2016_sps}, which is a normalized higher-order Lenard-Bernstein collisional operator~\cite{lenard_bernstein_1958_collisions}. The following hypercollisional operator is added to the right-hand side of Eq.~\eqref{f-vlasov-pde}:
\begin{equation}
    \mathcal{C}\left(C_{s, n}(x, t)\right) \coloneqq  -\nu \eta_{n}C_{s, n}(x, t)\qquad \mathrm{with} \qquad \eta_{n} \coloneqq \frac{n(n-1)(n-2)}{(N_{v}-1)(N_{v}-2)(N_{v}-3)},
\end{equation}
where the tunable parameter $\nu \in \mathbb{R}_{+}$ is the artificial collisional rate of the last Hermite coefficient $C_{s, N_{v}-1}(x, t)$. The artificial collisional operator does not act directly on the first three moments~($n=0, 1, 2$) of the Hermite expansion to maintain the conservation of mass, momentum, and energy~\cite{schumer_1998_sw_aw, camporeale_2016_sps}. In practice, the hypercollisional parameter 
$\nu \in \mathbb{R}_{+}$ is chosen as the minimal value that effectively damps unresolved fine-scale velocity structures that lead to recurrence or numerical instabilities while preserving the essential collisionless dynamics.

\subsection{Semi-Discrete Equations in Vector Form}\label{sec:vector_form}
The fluid-kinetic coupling property of the spectral method arises from the AW Hermite spectral expansion in velocity and is not influenced by the choice of spatial discretization. We discretize the spatial coordinate uniformly with $\Delta x$ mesh spacing and denote the discretized expansion coefficients as 
\begin{equation*}
    \mathbf{C}_{s, n}(t) \coloneqq [C_{s, n}(x_{1}, t), C_{s, n}(x_{2}, t), \ldots, C_{s, n}(x_{N_{x}}, t)]^{\top} \in \mathbb{R}^{N_{x}},
\end{equation*}
where $N_{x}$ is the number of mesh points in space and $x_{j}$ denotes the spatial grid location with index $j=1, 2, \ldots, N_{x}$. Discretizing Eqns.~\eqref{f-vlasov-pde}--\eqref{f-poisson-pde} using second-order centered finite differencing in space results in the following semi-discrete system of ordinary differential equations:
\begin{align}
    \frac{\mathrm{d}\mathbf{\Psi}_{s, F}(t)}{\mathrm{d}t} &= \mathbf{A}_{s, F}\mathbf{\Psi}_{s, F}(t) + \mathbf{B}_{s, F} \left[\mathbf{\Psi}_{s, F}(t) \otimes \mathbf{E}(t)\right] + \mathbf{G}_{s, F} \mathbf{\Psi}_{s, K}(t), \label{fluid-FOM-equations}\\
    \frac{\mathrm{d} \mathbf{\Psi}_{s, K}(t)}{\mathrm{d}t} &= \mathbf{A}_{s, K}\mathbf{\Psi}_{s, K}(t) + \mathbf{B}_{s, K} \left[\mathbf{\Psi}_{s, K}(t) \otimes \mathbf{E}(t)\right] + \mathbf{G}_{s, K}\mathbf{\Psi}_{s, F}(t) + \mathbf{J}_{s, K} \left[\mathbf{\Xi}_{F}^{\top}\mathbf{\Psi}_{s, F}(t) \odot \mathbf{E}(t)\right] , \label{kinetic-FOM-equations}\\
    \mathbf{D} \mathbf{E}(t) &= \sum_{s} q_{s} \alpha_{s} \mathbf{\Theta}_{F} \mathbf{\Psi}_{s, F}(t),\label{poisson-ode}
\end{align}
where $\otimes$ denotes the Kronecker product and $\odot$ denotes the Hadamard (element-wise) product. 
We refer to Eqns.~\eqref{fluid-FOM-equations}--\eqref{kinetic-FOM-equations} coupled with the semi-discrete Poisson equation~\eqref{poisson-ode} as the \textit{full-order model} (FOM). 
The fluid $\mathbf{\Psi}_{s, F}(t)$ and kinetic $\mathbf{\Psi}_{s, K}(t)$ state vectors are
\begin{alignat*}{3}
    \mathbf{\Psi}_{s, F}(t) &\coloneqq \left[\mathbf{C}_{s, 0}(t), \mathbf{C}_{s, 1}(t), \mathbf{C}_{s, 2}(t)\right]^{\top} \in  \mathbb{R}^{N_{F}} \qquad &&\mathrm{with} \qquad N_{F} \coloneqq 3N_{x}, \\
    \mathbf{\Psi}_{s, K}(t) &\coloneqq \left[\mathbf{C}_{s, 3}(t), \mathbf{C}_{s, 4}(t), \ldots,  \mathbf{C}_{s, N_{v}-1}(t)\right]^{\top} \in  \mathbb{R}^{N_{K}} \qquad &&\mathrm{with} \qquad N_{K} \coloneqq (N_{v}-3)N_{x}, 
\end{alignat*}
where the first three moments are the \textit{fluid} moments and the rest $N_{v}-3$ moments are the \textit{kinetic} moments. We are splitting the fluid and kinetic states because our objective is to apply POD exclusively to the kinetic part to be able to satisfy the conservation laws of mass, momentum, and energy.
The sparse matrices in the semi-discrete FOM Eqns.~\eqref{fluid-FOM-equations}--\eqref{poisson-ode} are defined as
\begin{align*}
    \mathbf{A}_{s, F} &\coloneqq \alpha_{s} \mathbf{A}_{1}(0, 3) + u_{s} \mathbf{A}_{2}(0, 3) + \nu \mathbf{A}_{3}(0, 3) \in \mathbb{R}^{N_{F} \times N_{F}},\\
    \mathbf{A}_{s, K} &\coloneqq \alpha_{s} \mathbf{A}_{1}(3, N_{v}) + u_{s} \mathbf{A}_{2}(3, N_{v}) + \nu \mathbf{A}_{3}(3, N_{v}) \in \mathbb{R}^{N_{K} \times N_{K}},\\
    \mathbf{B}_{s, F} &\coloneqq \frac{q_{s}}{m_{s}\alpha_{s}} \mathbf{B}(0, 3) \mathbf{Q}(0, 3)\in \mathbb{R}^{N_{F}\times N_{F}N_{x}}, \\
    \mathbf{B}_{s, K} &\coloneqq \frac{q_{s}}{m_{s}\alpha_{s}} \mathbf{B}(3, N_{v}) \mathbf{Q}(3, N_{v})\in \mathbb{R}^{N_{K}\times N_{K}N_{x}},\\
    \mathbf{\Theta}_{F} &\coloneqq [\mathbf{I}_{N_{x}}\Hquad \mathbf{0} \Hquad \ldots \Hquad \mathbf{0}]^{\top} \in \mathbb{R}^{N_{F} \times N_{x}}, \\
    \mathbf{\Xi}_{F} &\coloneqq [\mathbf{0} \Hquad \ldots \Hquad \mathbf{0} \Hquad \mathbf{I}_{N_{x}}]^{\top} \in \mathbb{R}^{N_{F} \times N_{x}}, \\
    \mathbf{G}_{s, F} &\coloneqq - \alpha_{s} \sigma_{3} \mathbf{\Xi}_{F} \mathbf{D} \mathbf{\Theta}_{K}^{\top}  = - \mathbf{G}_{s, K}^{\top} \in \mathbb{R}^{N_{F} \times N_{K}}, \\
    \mathbf{J}_{s, K} &\coloneqq \frac{2\sigma_{3} q_{s}}{m_{s}\alpha_{s}} \mathbf{\Theta}_{K} \in \mathbb{R}^{N_{K} \times N_{x}},
\end{align*}
where
\begin{align*}  
\mathbf{Q}(i, j) &\coloneqq \mathbf{I}_{(j-i)N_{x}} \bullet \left(\mathbf{1}_{j-i}^{\top} \otimes \mathbf{I}_{N_{x}}\right) \in \mathbb{R}^{N_{x}(j-i) \times N_{x}^2 (j-i)}, \\
\mathbf{A}_{1}(i, j) &\coloneqq -\begin{bmatrix}
    0 &  \sigma_{i+1}& 0 &  \ldots & 0 \\
    \sigma_{i+1} & 0 & \sigma_{i+2} &  \ldots & 0\\
    & &  & & \\
    & \ddots & \ddots &  \ddots&   \\
    & &  & & \\
    0 & 0 & \sigma_{j-2}  & 0 & \sigma_{j-1}   \\
    0 & 0 & 0 & \sigma_{j-1} & 0
 \end{bmatrix} \otimes \mathbf{D} \in \mathbb{R}^{N_{x} (j-i) \times N_{x}(j-i)}, \\
\mathbf{A}_{2}(i, j) &\coloneqq -\mathbf{I}_{j-i} \otimes \mathbf{D} \in \mathbb{R}^{N_{x} (j-i) \times N_{x}(j-i)}, \\
\mathbf{A}_{3}(i, j) &\coloneqq -\text{diag}(\eta_{i}, \eta_{i+1}, \ldots, \eta_{j-1}) \otimes \mathbf{I}_{N_{x}} \in \mathbb{R}^{N_{x} (j-i) \times N_{x}(j-i)}, \\
\mathbf{B}(i, j) &\coloneqq 2\begin{bmatrix}
    0 & 0 & 0 &  \ldots & 0 \\
    \sigma_{i+1} & 0 & 0 &  \ldots & 0\\
    & &  & & \\
    & \ddots & \ddots &  \ddots&   \\
    & &  & & \\
    0 & 0 & \sigma_{j-2} & 0 & 0\\
    0 & 0 & 0 & \sigma_{j-1} &0
 \end{bmatrix} \otimes \mathbf{I}_{N_{x}} \in \mathbb{R}^{N_{x}(j-i) \times N_{x}(j-i)}, \\
\mathbf{\Theta}_{K} &\coloneqq [\mathbf{I}_{N_{x}}\Hquad \mathbf{0} \Hquad \ldots \Hquad \mathbf{0}]^{\top} \in \mathbb{R}^{N_{K} \times N_{x}}, 
\end{align*}
such that $\mathbf{I}_{N}\in \mathbb{R}^{N \times N}$ is the identity matrix, $\mathbf{1}_{N} \in \mathbb{R}^{N}$ is a column vector of all ones,  and $\bullet$ is the face-splitting product (also known as the transposed Khatri-Rao product). The periodic second-order centered finite difference operator is defined as
\begin{equation}\label{central-finite-differencing}
    \mathbf{D} \coloneqq \frac{1}{2\Delta x}
    \left[
    \begin{array}{cccccc}
        0 & 1 & 0& & -1 \\
        \ddots & \ddots & \ddots & \ddots &\\
        & -1 & 0& 1 & 0 \\
         & & -1& 0 & 1 \\
        1 & & & -1 & 0 \\
    \end{array}
    \right]  \in \mathbb{R}^{N_{x} \times N_{x}}.
\end{equation}
%
\section{Conservative Model Order Reduction of Kinetic Moments via Proper Orthogonal Decomposition and Galerkin Projection}\label{sec:rom}
The proper orthogonal decomposition (POD) is a technique to compute the optimal basis for representing an experimental or simulated dataset by minimizing the mean squared error between the snapshot matrix (where each column is the simulation results at different timesteps and parameter realizations) and its reduced rank $N_{r}$ representation~\cite{berkooz_1993_pod, lumley_1967_structure}; see section~\ref{sec:pod} for further details.
POD requires performing a series of simulations \textit{a priori}, known as the \textit{training dataset}, by varying some of the tunable parameters.
Next, we apply Galerkin projection to the semi-discrete kinetic equations~\eqref{kinetic-FOM-equations}, projecting them onto the POD basis. 
We take advantage of the fluid-kinetic coupling property of the AW Hermite spectral expansion and only project the kinetic moments, thereby preserving mass, momentum, and energy conservation and retaining the fluid moments (macroscopic quantities) intact; see section~\ref{sec:reduced_equations}. 
In section~\ref{sec:affine_parametric}, we highlight that the spectral operators have an affine (linear) dependence on the model's tunable parameters, enabling efficient computation of the reduced operators for varying solver parameters (i.e., $\nu, \alpha_{s}, u_{s}$) and physical parameters (i.e., $q_{s}, m_{s}$). 
Lastly, we prove analytically in section~\ref{sec:conservation_properties} that the conservation properties are not impacted by projecting the kinetic moments.

\subsection{Proper Orthogonal Decomposition}\label{sec:pod}
Consider the snapshot matrix 
\begin{equation}\label{snapshot-matrix}
    \bar{\mathbf{\Psi}} \coloneqq \left[\mathbf{\Psi}_{s, K}(t_{1}; \mu_{1}), \ldots ,\mathbf{\Psi}_{s, K}(t_{N_{t}}; \mu_{1}), \ldots, \mathbf{\Psi}_{s, K}(t_{1}; \mu_{N_{\mu}}), \ldots, \mathbf{\Psi}_{s, K}(t_{N_{t}}; \mu_{N_{\mu}})\right] \in \mathbb{R}^{N_{K} \times N_{t}N_{\mu}},
\end{equation}
where $\mathbf{\Psi}_{s, K}(t; \mu) \in \mathbb{R}^{N_{K}}$ is the simulation kinetic state vector of dimensions $N_{K}$ at a given time $t$ and parameter realization $\mu$.
The parameter realization $\mu$ specifies the model parameters (i.e., $\nu, \alpha_{s}, u_{s}$) and physical parameters (i.e., $q_{s}, m_{s}$). 
We denote the total number of snapshots (for a single parameter realization) as $N_{t}$ and the total number of parametric realizations as $N_{\mu}$. The POD basis $\{\mathbf{p}_{1}, \ldots, \mathbf{p}_{N_{r}}\}$ satisfies the following optimization problem:
\begin{equation*}
    \min_{\mathbf{p}_{1}, \ldots, \mathbf{p}_{N_{r}}} \left\lVert \sum_{j=1}^{N_{t}N_{\mu}} \left[\bar{\mathbf{\Psi}}_{j} - \sum_{i=1}^{N_{r}} \left(\bar{\mathbf{\Psi}}_{j}^{\top} \mathbf{p}_{i} \right) \mathbf{p}_{i} \right]\right\rVert_{2}^{2} \qquad \mathrm{s.t.} \qquad \mathbf{p}_{i}^{\top} \mathbf{p}_{j} = \delta_{ij} \qquad \mathrm{and} \qquad N_{r} \ll N_{K},
\end{equation*}
where $\delta_{i,j}$ is the Kronecker delta function, $\bar{\mathbf{\Psi}}_{i} \in \mathbb{R}^{N_{K}}$ is the $i$th column of the snapshot matrix $\bar{\mathbf{\Psi}} \in \mathbb{R}^{N_{K} \times N_{t}N_{\mu}}$ defined in Eq.~\eqref{snapshot-matrix}, and $N_{r}$ is the number of POD basis vectors used to approximate the problem of size $N_{K}$. 
Therefore, the POD basis offers a reduced set of basis vectors that minimizes the mean squared error between the simulation dataset and its rank $N_{r}$ approximation. 
In principle, the POD basis $\{\mathbf{p}_{1}, \ldots, \mathbf{p}_{N_{r}}\}$ is computed by an eigenvalue decomposition of the matrix $\mathbf{\bar{\Psi}} \mathbf{\bar{\Psi}}^{\top} \in \mathbb{R}^{N_{K} \times N_{K}}$, or more efficiently when $N_{K} \gg N_{t}N_\mu$ via the \textit{method of snapshots}~\cite{sirovich_1987_svd} which uses $\mathbf{\bar{\Psi}}^\top \mathbf{\bar{\Psi}} \in \mathbb{R}^{N_{t}N_\mu \times N_{t}N_\mu}$, or also directly via the singular value decomposition of the snapshot matrix $\bar{\mathbf{\Psi}} = \mathbf{P} \mathbf{S} \mathbf{V}^{\top}$, such that $\mathbf{P} \in \mathbb{R}^{N_{K} \times N_{K}}$ and $\mathbf{V} \in \mathbb{R}^{N_{t}N_{\mu} \times N_{t} N_{\mu}}$ are unitary matrices and $\mathbf{S} \in \mathbb{R}^{N_{K} \times N_{t}N_{\mu}}$ is a diagonal matrix with the singular values $\sigma_{i}$ arranged in decreasing order on its diagonal. The POD basis is the first $N_{r}$ columns of the left singular vectors matrix $\mathbf{P}\in \mathbb{R}^{N_{K} \times N_{K}}$, which we denote by $\hat{\mathbf{P}} \in \mathbb{R}^{N_{K} \times N_{r}}$. The number of reduced dimensions $N_{r}$ is typically chosen based on where the singular values plateau or reach a certain threshold. 

\subsection{Galerkin Projection for ROM Construction of the Kinetic Part of the Spectral Expansion}\label{sec:reduced_equations}
We assume there is a low-dimensional representation of the kinetic state vector $\mathbf{\Psi}_{s, K}(t) \in \mathbb{R}^{N_{K}}$, such that
\begin{equation*}
    \mathbf{\Psi}_{s, K}(t) \approx \hat{\mathbf{P}} \hat{\mathbf{\Psi}}_{s, K}(t) \qquad \text{s.t.} \qquad \hat{\mathbf{P}}^{\top} \hat{\mathbf{P}} = \mathbf{I}_{N_{r}}\qquad \mathrm{and} \qquad \qquad N_{r} \ll N_{K},
\end{equation*}
where $\hat{\mathbf{\Psi}}_{s, K}(t)\in \mathbb{R}^{N_{r}}$ and $\hat{\mathbf{P}} \in \mathbb{R}^{N_{K}\times N_{r}}$ are the reduced kinetic state and POD basis of species $s$, respectively.
The Galerkin projection of Eq.~\eqref{kinetic-FOM-equations} yields the following reduced equation
\begin{align}\label{vector-vlasov-rom-1}
   \frac{\mathrm{d} \hat{\mathbf{\Psi}}_{s, K}(t)}{\mathrm{d} t} = \hat{\mathbf{P}}^{\top}\mathbf{A}_{s, K} \hat{\mathbf{P}} \hat{\mathbf{\Psi}}_{s, K}(t) &+ \underbrace{\hat{\mathbf{P}}^{\top} \mathbf{B}_{s, K}}_{N_{r} \times N_{K}N_{x}} \underbrace{\left[\hat{\mathbf{P}} \hat{\mathbf{\Psi}}_{s, K}(t) \otimes \mathbf{E}(t)\right]}_{N_{K}N_{x}} + \hat{\mathbf{P}}^{\top} \mathbf{G}_{s, K}\mathbf{\Psi}_{s, F}(t) \\ 
   &+\hat{\mathbf{P}}^{\top} \mathbf{J}_{s, K} \left[\mathbf{\Xi}_{F}^{\top}\mathbf{\Psi}_{s, F}(t) \odot \mathbf{E}(t)\right]. \nonumber
\end{align}
Currently, evolving Eq.~\eqref{vector-vlasov-rom-1} computationally scales with the high-dimensional kinetic discretization $N_{K}$ instead of the low-dimensional $N_{r}$ due to the nonlinear acceleration term. This is often referred to as the \textit{nonlinear bottleneck} in model reduction, where the linear low-dimensional projection of nonlinear high-dimensional equations still scales computationally with the dimensions of the original high-dimensional equations. We overcome the nonlinear bottleneck via a simple Kronecker product identity: 
\begin{equation*}
    \hat{\mathbf{P}}\hat{\mathbf{\Psi}}_{s, K}(t) \otimes \mathbf{E}(t) = \left[\hat{\mathbf{P}}\otimes \mathbf{I}_{N_{x}} \right]\left[ \hat{\mathbf{\Psi}}_{s, K}(t) \otimes \mathbf{E}(t)\right].
\end{equation*}
By employing the identity above and the Kronecker identity to simplify the nonlinearity in Eq.~\eqref{vector-vlasov-rom-1}, we get
\begin{equation*}
    \frac{\mathrm{d}\hat{\mathbf{\Psi}}_{s, K}(t)}{\mathrm{d} t}  = \hat{\mathbf{A}}_{s, K} \hat{\mathbf{\Psi}}_{s, K}(t) + \underbrace{\hat{\mathbf{B}}_{s, K}}_{N_{r} \times N_{r} N_{x}} \underbrace{\left[\hat{\mathbf{\Psi}}_{s, K}(t) \otimes \mathbf{E}(t)\right]}_{N_{r}N_{x}}+ \hat{\mathbf{G}}_{s, K}\mathbf{\Psi}_{s, F}(t) + \hat{\mathbf{J}}_{s, K}\left[\mathbf{\Xi}_{F}^{\top}\mathbf{\Psi}_{s, F}(t) \odot \mathbf{E}(t)\right].
\end{equation*}
Therefore, evolving the reduced kinetic equation scales with the reduced dimension $N_{r}\ll N_{K}$, 
where the reduced advection matrix is $\hat{\mathbf{A}}_{s, K} \coloneqq \hat{\mathbf{P}}^{\top}\mathbf{A}_{s,  K} \mathbf{P} \in \mathbb{R}^{N_{r} \times N_{r}}$, the reduced acceleration matrix is 
$\hat{\mathbf{B}}_{s, K} \coloneqq  \hat{\mathbf{P}}^{\top} \mathbf{B}_{s, K}[\hat{\mathbf{P}} \otimes \mathbf{I}_{N_{x}}]\in \mathbb{R}^{N_{r} \times N_{r}N_{x}}$,
and the fluid-kinetic coupling reduced matrices are $\hat{\mathbf{G}}_{s, K}  \coloneqq \hat{\mathbf{P}}^{\top} \mathbf{G}_{s, K} \in\mathbb{R}^{N_{r} \times N_{F}}$ and $\hat{\mathbf{J}}_{s, K} \coloneqq \hat{\mathbf{P}}^{\top} \mathbf{J}_{s, K} \in\mathbb{R}^{N_{r} \times N_{x}}$. To summarize, the semi-discrete \textit{reduced-order model} (ROM) equations are
\begin{align}
   \frac{\mathrm{d} \mathbf{\Psi}_{s, F}(t)}{\mathrm{d}t} &= \mathbf{A}_{s, F} \mathbf{\Psi}_{s, F}(t) + \mathbf{B}_{s, F} \left[\mathbf{\Psi}_{s, F}(t)\otimes \mathbf{E}(t)\right] + \hat{\mathbf{G}}_{s, F} \hat{\mathbf{\Psi}}_{s, K}(t), \label{fluid-ROM-equations}\\
    \frac{\mathrm{d}\hat{\mathbf{\Psi}}_{s, K}(t)}{\mathrm{d} t}  &= \hat{\mathbf{A}}_{s, K}\hat{\mathbf{\Psi}}_{s, K}(t) + \hat{\mathbf{B}}_{s, K} \left[\hat{\mathbf{\Psi}}_{s, K}(t) \otimes \mathbf{E}(t) \right]+ \hat{\mathbf{G}}_{s, K}\mathbf{\Psi}_{s, F}(t) + \hat{\mathbf{J}}_{s, K}\left[\mathbf{\Xi}_{F}^{\top}\mathbf{\Psi}_{s, F}(t) \odot \mathbf{E}(t)\right], \label{kinetic-ROM-equations}
\end{align}
coupled with the semi-discrete Poisson equation~\eqref{poisson-ode} and with $\hat{\mathbf{G}}_{s, F} \coloneqq \mathbf{G}_{s, F} \hat{\mathbf{P}} \in \mathbb{R}^{N_{F} \times N_{r}}$. While the overall structure of the equations remains unchanged, the FOM~\eqref{kinetic-FOM-equations} consists of high-dimensional sparse operations evolving $N_K$ DOFs, whereas the ROM~\eqref{kinetic-ROM-equations} consists of low-dimensional dense operations evolving $N_r \ll N_K$ DOFs.

\subsection{Affine Parametric Dependence}\label{sec:affine_parametric}
The kinetic FOM Eq.~\eqref{kinetic-FOM-equations} depends on a set of tunable model parameters: the Hermite parameters $u_{s} \in \mathbb{R}$ and $\alpha_{s} \in \mathbb{R}_{+}$, and the artificial collisional parameter $\nu \in \mathbb{R}_{+}$. The same holds for mass $m_{s}$ and charge $q_{s}$. The kinetic FOM operators in Eq.~\eqref{kinetic-FOM-equations}, denoted by $\mathbf{A}_{s, K}, \mathbf{B}_{s, K}, \mathbf{G}_{s, K},$ and $\mathbf{J}_{s, K}$, all have the simplest form of affine-parametric dependence on the model parameters such that $\mathbf{A}(\mu) = \mu \mathbf{A}$, where $\mu$ is the parameter and $\mathbf{A}$ is the operator; see section~\ref{sec:reduced_equations} for the kinetic FOM operators' definition. 
This structure allows us to efficiently construct the parametric reduced operators $\hat{\mathbf{A}}_{s, K}, \hat{\mathbf{B}}_{s, K}, \hat{\mathbf{G}}_{s, K},$ and $\hat{\mathbf{J}}_{s, K}$, such that changing the model parameters does not require re-projecting the kinetic FOM operators onto the POD basis. This is crucial for practical purposes since, due to the affine parametric dependence, the construction of reduced operators is a one-time \textit{a priori} computation.

\subsection{Conservation Properties: Mass, Momentum, and Energy}\label{sec:conservation_properties}
The fluid FOM Eq.~\eqref{fluid-FOM-equations} is identical to the fluid ROM Eq.~\eqref{fluid-ROM-equations} besides the coupling term to the kinetic state. However, as shown below, the difference in the fluid-kinetic coupling term does not influence the ROM's conservation of mass, momentum, and energy. 

\subsubsection{Mass Conservation}
The mass of species $s$ is defined as 
\begin{equation*}
    \mathcal{M}_{s}(t) \coloneqq \int_{0}^{\ell} \int_{\mathbb{R}} f_{s}(x, v, t)\mathrm{d} v \mathrm{d} x = \Delta x \alpha_{s} \| \mathbf{C}_{s, 0}(t)\|_{1}.
\end{equation*}
The FOM~\eqref{fluid-FOM-equations} and ROM~\eqref{fluid-ROM-equations} evolution of $\mathbf{C}_{s, 0}(t)$ are identical and lead to 
\begin{equation*}
    \frac{\mathrm{d}\mathbf{C}_{s, 0}(t)}{\mathrm{d} t} = - \mathbf{D} \left[\frac{\alpha_{s}}{\sqrt{2}}\mathbf{C}_{s, 1}(t) + u_{s} \mathbf{C}_{s, 0}(t)\right].
\end{equation*}
Since the central finite difference derivative operator $\mathbf{D} \in \mathbb{R}^{N_{x} \times N_{x}}$ defined in Eq.~\eqref{central-finite-differencing} is skew-symmetric, i.e. $\mathbf{D} = - \mathbf{D}^{\top}$, we get $\mathbf{1}^{\top} \mathbf{D} = [\mathbf{D}^{\top} \mathbf{1}]^{\top} =- [\mathbf{D} \mathbf{1}]^{\top} = \mathbf{0}^{\top}$,  where $\mathbf{1} \in \mathbb{R}^{N_{x}}$  is a column vector of all ones and $\mathbf{0} \in \mathbb{R}^{N_{x}}$ is a column vector of all zeros. Thus, 
\begin{equation*}
  \frac{\mathrm{d}\|\mathbf{C}_{s, 0}(t)\|_{1}}{\mathrm{d} t} 
 = \mathbf{1}^{\top} \frac{\mathrm{d}\mathbf{C}_{s, 0}(t)}{\mathrm{d} t} = - \cancelto{\mathbf{0}^{\top}}{\mathbf{1}^{\top} \mathbf{D}} \left[\frac{\alpha_{s}}{\sqrt{2}} \mathbf{C}_{s, 1}(t) + u_{s} \mathbf{C}_{s, 0}(t)\right]  = 0 \qquad \Rightarrow \qquad \frac{\mathrm{d} \mathcal{M}_{s}(t)}{\mathrm{d} t} = 0.
\end{equation*}
%

\subsubsection{Momentum Conservation}
The total momentum is defined as 
\begin{equation*}
    \mathcal{P}(t) \coloneqq \sum_{s} m_{s} \int_{0}^{\ell} \int_{\mathbb{R}} v f_{s}(x, v, t)\mathrm{d} v \mathrm{d} x = \Delta x \sum_{s} \alpha_{s} m_{s} \left[ \frac{\alpha_{s}}{\sqrt{2}} \|\mathbf{C}_{s, 1}(t) \|_{1} + u_{s} \|\mathbf{C}_{s, 0}(t)\|_{1}\right],
\end{equation*}
The time derivative of the total momentum is 
\begin{equation}\label{change-in-momentum}
    \frac{\mathrm{d}\mathcal{P}(t)}{\mathrm{d}t} = \Delta x \sum_{s} m_{s} \alpha_{s} \left[\frac{\alpha_{s}}{\sqrt{2}} \frac{\mathrm{d} \|\mathbf{C}_{s, 1}(t)\|_{1}}{\mathrm{d} t} + u_{s} \cancelto{0}{\frac{\mathrm{d} \|\mathbf{C}_{s, 0}(t)\|_{1}}{\mathrm{d} t}}\right].
\end{equation}
The FOM~\eqref{fluid-FOM-equations} and ROM~\eqref{fluid-ROM-equations} evolution of $\mathbf{C}_{s, 1}(t)$ are identical and lead to 
\begin{align}
    \frac{\mathrm{d} \|\mathbf{C}_{s, 1}(t)\|_{1}}{\mathrm{d} t}  &= \mathbf{1}^{\top} \frac{\mathrm{d}\mathbf{C}_{s, 1}(t)}{\mathrm{d} t} 
    = - \cancelto{\mathbf{0}^{\top}}{\mathbf{1}^{\top}\mathbf{D}}\left[ \frac{\alpha_{s}}{\sqrt{2}}  \mathbf{C}_{s, 0}(t) + \alpha_{s} \mathbf{C}_{s, 2}(t) + u_{s} \mathbf{C}_{s, 1}(t) \right] + \frac{\sqrt{2}q_{s}}{\alpha_{s} m_{s}} \mathbf{1}^{\top} \left[\mathbf{C}_{s, 0}(t) \odot \mathbf{E}(t)\right]\label{change-in-c1-momentum}\\
    &= \frac{\sqrt{2}q_{s}}{\alpha_{s} m_{s}} \mathbf{E}(t)^{\top} \mathbf{C}_{s, 0}(t)\nonumber.
\end{align}
Inserting Eq.~\eqref{change-in-c1-momentum} in Eq.~\eqref{change-in-momentum} and employing the semi-discrete Poisson equation~\eqref{poisson-ode} results in
\begin{equation*}
    \frac{\mathrm{d}\mathcal{P}(t)}{\mathrm{d} t} =  \Delta x \mathbf{E}(t)^{\top}  \underbrace{\sum_{s} \alpha_{s}q_{s}\mathbf{C}_{s, 0}(t)}_{\text{insert Eq.}~\eqref{poisson-ode}} =\Delta x  \mathbf{E}(t)^{\top} \mathbf{D} \mathbf{E}(t) = 0.
\end{equation*}
In the above, we use the skew-symmetric property of $\mathbf{D}$, i.e. $\mathbf{D} = - \mathbf{D}^{\top}$, which leads to $\mathbf{E}(t)^{\top} \mathbf{D} \mathbf{E}(t) = [\mathbf{E}(t)^{\top} \mathbf{D}\mathbf{E}(t)]^{\top} = - \mathbf{E}(t)^{\top} \mathbf{D} \mathbf{E}(t) = 0$.
Thus, total momentum is conserved at the semi-discrete level.

\subsubsection{Energy Conservation}
The total energy is the sum of kinetic and potential energies denoted as $\mathcal{E}(t) \coloneqq \mathcal{E}_{\mathrm{kin}}(t) + \mathcal{E}_{\mathrm{pot}}(t)$, where
\begin{align*}
    \mathcal{E}_{\mathrm{kin}}(t) &\coloneqq \sum_{s} \frac{m_{s}}{2} \int_{0}^{\ell} \int_{\mathbb{R}} v^{2} f_{s}(x, v, t)\mathrm{d} v \mathrm{d} x \\
    &= \frac{\Delta x}{2}\sum_{s} \alpha_{s} m_{s}\left[\frac{\alpha_{s}^{2}}{\sqrt{2}} \|\mathbf{C}_{s, 2}(t)\|_{1}+ \sqrt{2} u_{s} \alpha_{s} \| \mathbf{C}_{s, 1}(t) \|_{1}+ \left(\frac{\alpha_{s}^2}{2} + u_{s}^{2}\right) \| \mathbf{C}_{s, 0}(t) \|_{1}\right]
\end{align*}
and
\begin{equation}\label{potential-energy-definition}
    \mathcal{E}_{\mathrm{pot}}(t) \coloneqq \frac{1}{2} \int_{0}^{\ell}E(x, t)^2 \mathrm{d} x = \frac{\Delta x}{2} \mathbf{E}(t)^{\top} \mathbf{E}(t) 
\end{equation}
The time derivative of the kinetic energy is 
\begin{align}
    \frac{\mathrm{d} \mathcal{E}_{\mathrm{kin}}(t)}{\mathrm{d} t} = \frac{\Delta x}{2}\sum_{s} \alpha_{s} m_{s}\left[\frac{\alpha_{s}^{2}}{\sqrt{2}} \frac{\mathrm{d} \|\mathbf{C}_{s, 2}(t)\|_{1}}{\mathrm{d} t} + \sqrt{2} u_{s} \alpha_{s}\frac{\mathrm{d} \| \mathbf{C}_{s, 1}(t) \|_{1}}{\mathrm{d} t} + \left(\frac{\alpha_{s}^2}{2} + u_{s}^{2}\right) \cancelto{0}{\frac{\mathrm{d}\| \mathbf{C}_{s, 0}(t) \|_{1}}{\mathrm{d} t}}\right]\label{change-in-kinetic-energy}
\end{align}
Although the FOM~\eqref{fluid-FOM-equations} and ROM~\eqref{fluid-ROM-equations} describe different evolutions for $\mathbf{C}_{s, 2}(t)$, particularly in the term coupling to $\mathbf{C}_{s, 3}(t)$, due to the skew-symmetric property of $\mathbf{D}$, both equations lead to 
\begin{equation}\label{c2-evolution}
    \frac{\mathrm{d} \|\mathbf{C}_{s, 2}(t)\|_{1}}{\mathrm{d} t} = \frac{2q_{s}}{\alpha_{s} m_{s}} \mathbf{E}(t)^{\top} \mathbf{C}_{s, 1}(t).
\end{equation}
Inserting Eq.~\eqref{c2-evolution} and Eq.~\eqref{change-in-c1-momentum} in Eq.~\eqref{change-in-kinetic-energy} results in 
\begin{equation}\label{change-in-kinetic-energy-2}
     \frac{\mathrm{d} \mathcal{E}_{\mathrm{kin}}(t)}{\mathrm{d} t} = \Delta x \mathbf{E}(t)^{\top} \sum_{s} \alpha_{s} q_{s} \left[ \frac{\alpha_{s}}{\sqrt{2}} \mathbf{C}_{s, 1}(t) + u_{s} \mathbf{C}_{s, 0}(t) \right].
\end{equation}
We can derive the Ampere equation from simple algebraic manipulation of the semi-discrete Poisson equation~\eqref{poisson-ode}, such that
\begin{align}
    \mathbf{D} \frac{\mathrm{d} \mathbf{E}(t)}{\mathrm{d} t} &= \sum_{s} q_{s} \alpha_{s} \frac{\mathrm{d} \mathbf{C}_{s, 0}(t)}{\mathrm{d} t} = -\mathbf{D} \sum_{s} q_{s} \alpha_{s} \left[\frac{\alpha_{s}}{\sqrt{2}}\mathbf{C}_{s, 1}(t) + u_{s} \mathbf{C}_{s, 0}(t) \right]\nonumber\\ 
    \frac{\mathrm{d} \mathbf{E}(t)}{\mathrm{d} t} &= -\sum_{s} q_{s} \alpha_{s} \left[\frac{\alpha_{s}}{\sqrt{2}}\mathbf{C}_{s, 1}(t) + u_{s} \mathbf{C}_{s, 0}(t) \right] + c \mathbf{1},\label{poisson-manipulated}
\end{align}
where $c \in \mathbb{R}$ is an arbitrary constant. Inserting the semi-discrete Ampere equation~\eqref{poisson-manipulated} in Eq.~\eqref{change-in-kinetic-energy-2} and leveraging the definition of the potential energy in Eq.~\eqref{potential-energy-definition} results in
\begin{equation*}
    \frac{\mathrm{d} \mathcal{E}_{\mathrm{kin}}(t)}{\mathrm{d} t} = -\Delta x \mathbf{E}(t)^{\top}  \frac{\mathrm{d} \mathbf{E}(t)}{\mathrm{d} t} = -\frac{\mathrm{d} \mathcal{E}_{\mathrm{pot}}(t)}{\mathrm{d} t} \qquad \Longrightarrow \qquad \frac{\mathrm{d} \mathcal{E}}{\mathrm{d} t} = 0.
\end{equation*}  
%

\section{Numerical Results}\label{sec:numerical-results}
In this section, we test the parametric ROM on two electrostatic benchmark problems. The temporal integrator, initial condition, and parametric setup are described in section~\ref{sec:implementation-details}. The numerical results for the weak Landau damping problem are presented in section~\ref{sec:weak-landau-damping-results}, and the two-stream instability in section~\ref{sec:two-stream-instability-results}. 

\subsection{Implementation Details}\label{sec:implementation-details}
We use the second-order implicit midpoint temporal integrator since it conserves linear and quadratic invariants of the semi-discrete system~\cite{hairer_2006_geometric, pagliantini_2023_rk}, i.e., mass, momentum, and energy. The nonlinear system~\eqref{fluid-FOM-equations}--\eqref{poisson-ode} at each time step is solved using an unpreconditioned Jacobian-Free-Newton-Krylov (JFNK) method~\cite{knoll_jfnk_2004} with absolute tolerance set to $10^{-12}$ and the time interval is $[0, T]$. The linear Krylov solver that is embedded in the JFNK method is the linear generalized minimum residual (LGMRES) method~\cite{baker_2005_lgmres} with relative and absolute tolerance set to $10^{-5}$. We solve the semi-discrete Poisson equation~\eqref{poisson-ode} via the LGMRES method at each time step with relative and absolute tolerance set to $10^{-12}$. The timestep is set to $\Delta t = 10^{-2}$.
All simulations are performed on a MacBook Pro 2.3 GHz Quad-Core Intel Core i7 processor with 16 GB RAM.

We initialize the particle distribution function of species $s$ as perturbed Maxwellians 
\begin{equation*}
    f_{s}(x, v, t=0) = \frac{n_{s} (1 + \epsilon \cos(x))}{\sqrt{\pi} \alpha_{s}}\exp\left(-\frac{\left(v-u_{s}\right)^2}{\alpha_{s}^2}\right) \qquad \Rightarrow \qquad C_{s, 0}(x, t=0) = \frac{n_{s} (1 + \epsilon \cos(x))}{\alpha_{s}},
\end{equation*}
where $\epsilon$ is the amplitude of the initial perturbation, $n_{s}$ is the initial average density, and $u_{s}$ and $\alpha_{s}$ are the Hermite velocity shifting and scaling parameters, respectively. The FOM and ROM initial conditions are the same since the only nonzero coefficient at $t=0$ is the zeroth Hermite coefficient. We consider the ions as a stationary neutralizing background, which appears as a source term in the Poisson equation. 
The setup parameters for the weak Landau damping and two-stream instability are listed in Table~\ref{tab:setup-parameters}. For each benchmark problem, we vary a single parameter and use four uniformly spaced samples for training. In high-dimensional parameter spaces, more strategic sample selection can be achieved through adaptive methods such as greedy search or local sensitivity analysis~\cite{benner_2015_parametric_survey}.

\noindent
\begin{table}
\centering
\caption{\label{tab:setup-parameters} Numerical simulation parameter setup for the FOM training dataset: weak Landau damping and two-stream instability. }
\setlength{\tabcolsep}{0.5em} 
\renewcommand{\arraystretch}{1.1} 
\begin{tabular}{p{2cm}|p{4.5cm}|p{6cm}}
parameters & weak Landau damping & two-stream instability \\
\hline
$s$ & electrons and immobile ions $s=e$& two electron beams and immobile ions $s\in\{e_{2}, e_{1}\}$\\
\hline
$N_{x}$ & $151$ & $251$ \\
\hline
$N_{v}$ & $50$ & $350$ \\
\hline
$\ell$ & $2\pi$ & $2\pi$ \\
\hline
$\nu$ & $10$ & $15$ \\
\hline
$\epsilon$ & $0.01$ & $ 0.1$ \\
\hline
$T$ & $10$ & $30$ \\
\hline 
$n_{s}$ & $n_{e} = 1$ & $n_{e_{1}} = n_{e_{2}} = 0.5$ \\
\hline 
$u_{s}$ & $u_{e} = 0$ & $u_{e_{2}} = -u_{e_{1}} = \{1.05, 1.06, 1.07, 1.08\}$  \\
\hline 
$\alpha_{s}$ & $\alpha_{e}=\{0.6, 0.7, 0.8, 0.9\}$ & $\alpha_{e_{1}} = \alpha_{e_{2}} 
 = 0.5$ \\
\end{tabular}
\end{table}

\subsection{Weak Landau Damping} \label{sec:weak-landau-damping-results}
\begin{figure}
    \centering
        \begin{subfigure}[b]{0.305\textwidth}
        \centering 
        \caption{Weak Landau initial condition}
        \label{fig:weak_landau_initial_condition}
        \includegraphics[width=\textwidth]{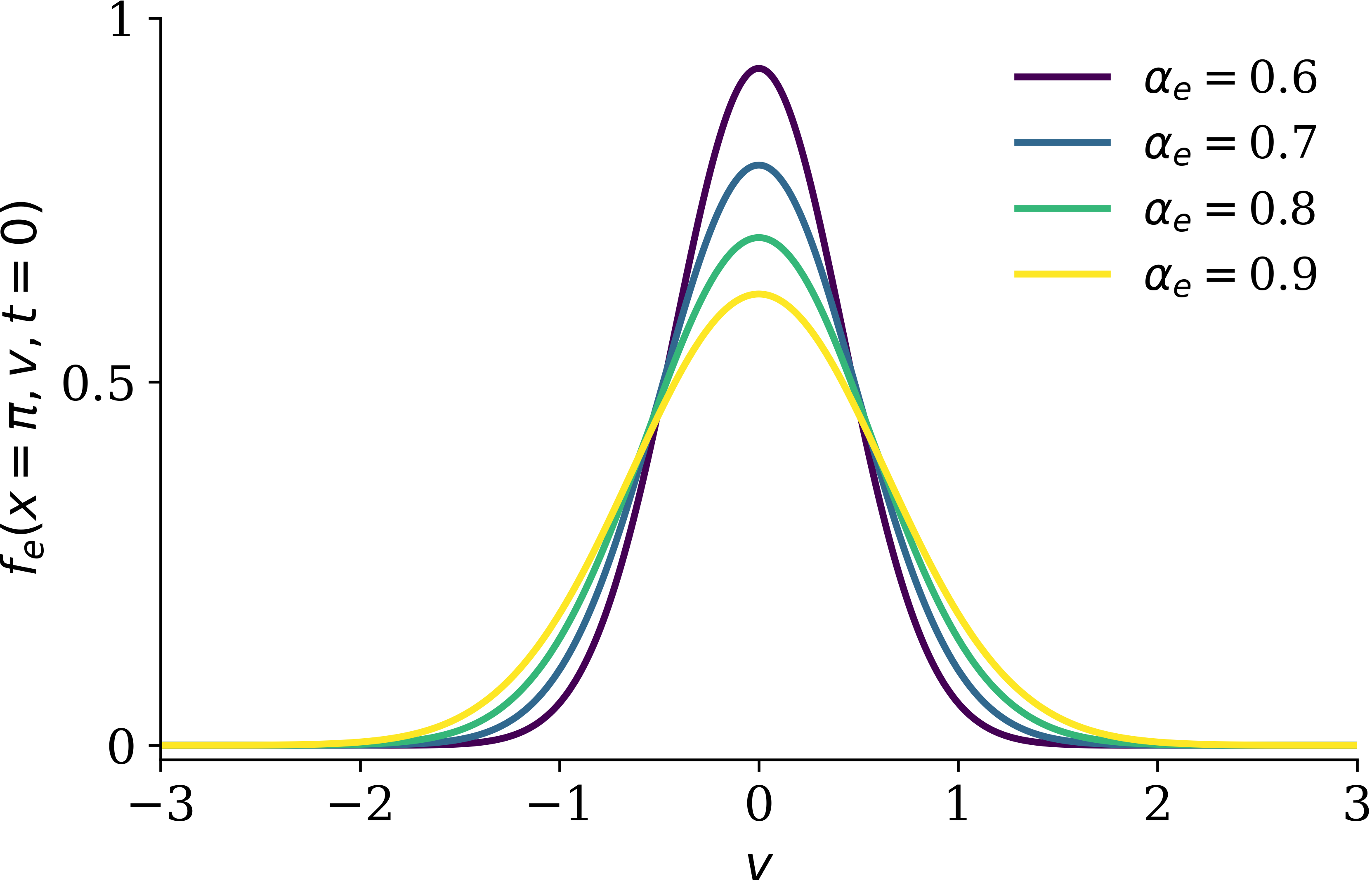}
    \end{subfigure}
    \hspace{-10pt}
    \begin{subfigure}[b]{0.39\textwidth}
        \centering 
        \caption{Weak Landau electric field first Fourier mode magnitude}
        \label{fig:weak_landau_electric_field_amplitude}
        \includegraphics[width=\textwidth]{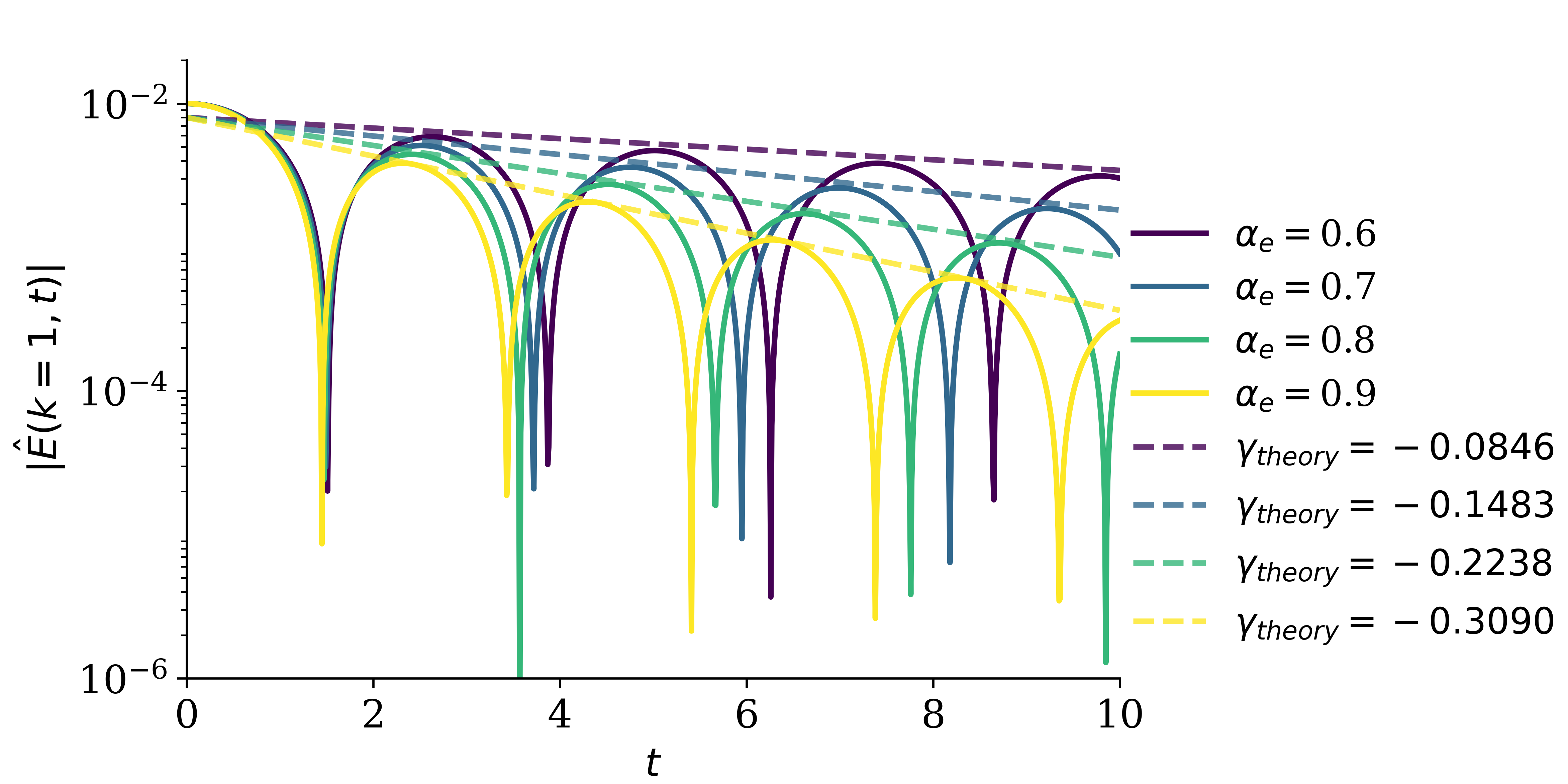}
    \end{subfigure}
    \hspace{-6pt}
    \begin{subfigure}[b]{0.305\textwidth}
        \centering 
        \caption{Weak Landau singular value decay}
        \label{fig:weak_landau_singular_value_decay}        \includegraphics[width=\textwidth]{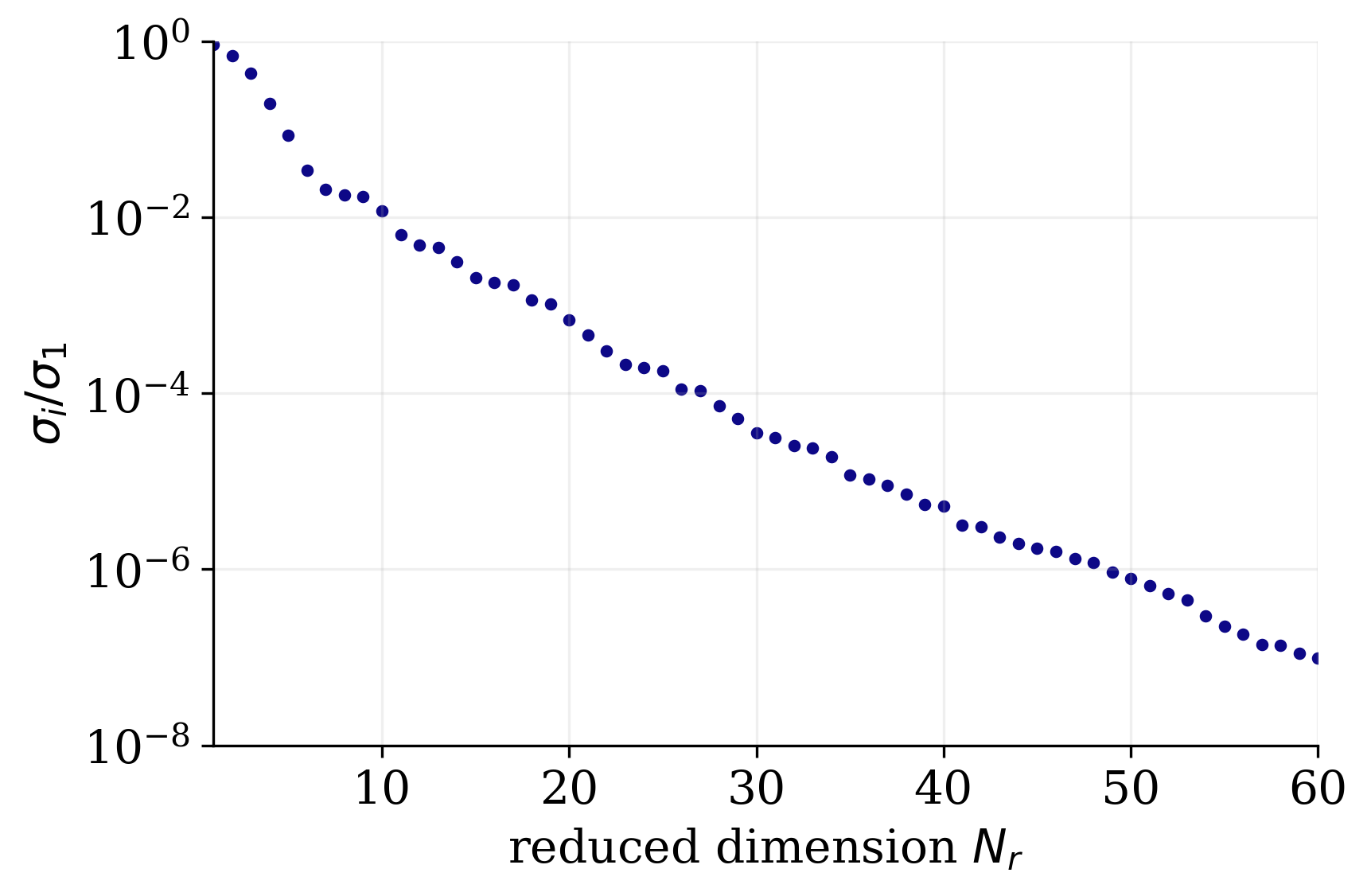}
    \end{subfigure}
    \caption{Weak Landau damping simulation data used to construct the POD basis. We vary the initial condition thermal velocity, see subfigure~(a), which alters the damping rate of the electric field, see subfigure~(b). Subfigure~(c) shows the normalized singular value decay, indicating that a low-dimensional representation exists since the FOM kinetic state dimensions are $N_{K} = (N_{v}-3)N_{x} = 7,097$. }
    \label{fig:weak_landau_damping_setup}
\end{figure}

\begin{figure}
    \centering
    \includegraphics[width=\textwidth]{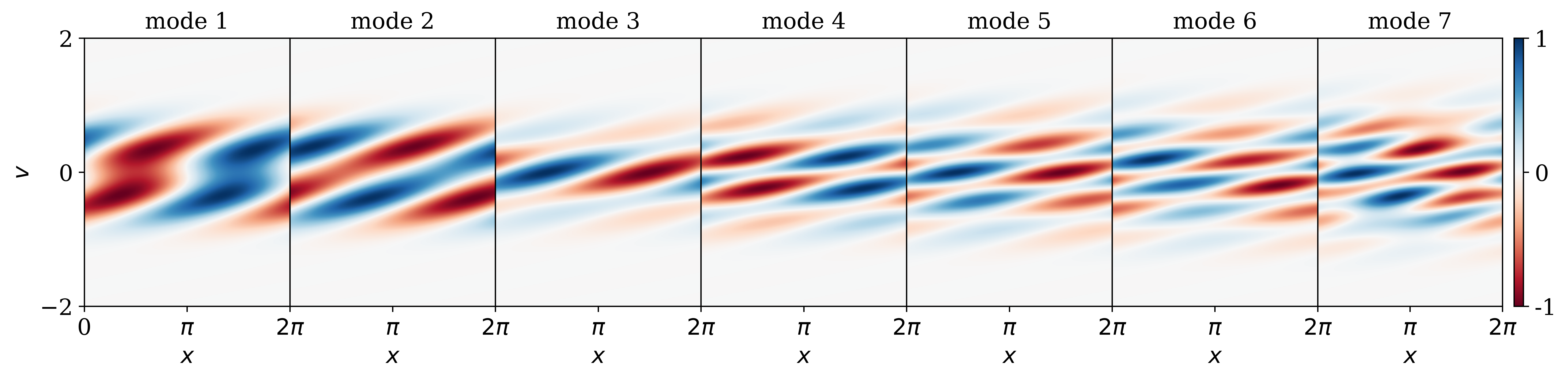}
    \caption{Weak Landau damping first seven POD modes in phase space (normalized to respective maximum value). The modes represent structures caused by phase space mixing. The higher modes capture small-scale velocity structures and filamentation. }
    \label{fig:weak_landau_damping_pod_modes}
\end{figure}

We investigate the ROM's ability to predict the weak Landau damping benchmark problem~\cite{landau_46}.
We vary the electron thermal velocity $\alpha_{e} = \{0.6, 0.7, 0.8, 0.9\}$ for this parametric study. 
Figure~\ref{fig:weak_landau_initial_condition} shows the training initial conditions varying the electron thermal velocity $\alpha_{e}$, and Figure~\ref{fig:weak_landau_electric_field_amplitude} shows the corresponding FOM electric field first Fourier mode magnitude $|\hat{E}(k=1, t)|$, along with analytic damping rates $\gamma_{\mathrm{theory}} \in \mathbb{R}$ by solving the linear dispersion relation. 
We simulate the FOM  for $t \in [0, 10]$ time interval, then we append all four parametric solutions into the snapshot matrix and perform the singular value decomposition to compute the POD basis. The singular value decay of the snapshot matrix is shown in Figure~\ref{fig:weak_landau_singular_value_decay}, which indicates that a low-dimensional linear subspace approximation is possible, as the kinetic state vector is of dimension $N_{K} = 7,097$ and the fast singular value decay leads to $\sigma_{50}/\sigma_{1} \sim 10^{-6}$. 
Note that in this problem, as well as in the following two-stream instability problem, $99.99\%$ of the total energy of the system is retained within the first five modes, but the ROM requires a significantly larger number of modes for reasonable accuracy in representing the dynamics. 
Figure~\ref{fig:weak_landau_damping_pod_modes} shows the first seven POD modes normalized to their respective absolute maximum values. The higher POD modes capture more small-scale velocity structures, in which their superposition describes phase-space mixing and filamentation. 

\begin{figure}
    \centering
    \begin{subfigure}[b]{0.5\textwidth}
        \centering 
        \caption{$\alpha_{e} = 0.75$}
        \includegraphics[width=\textwidth]{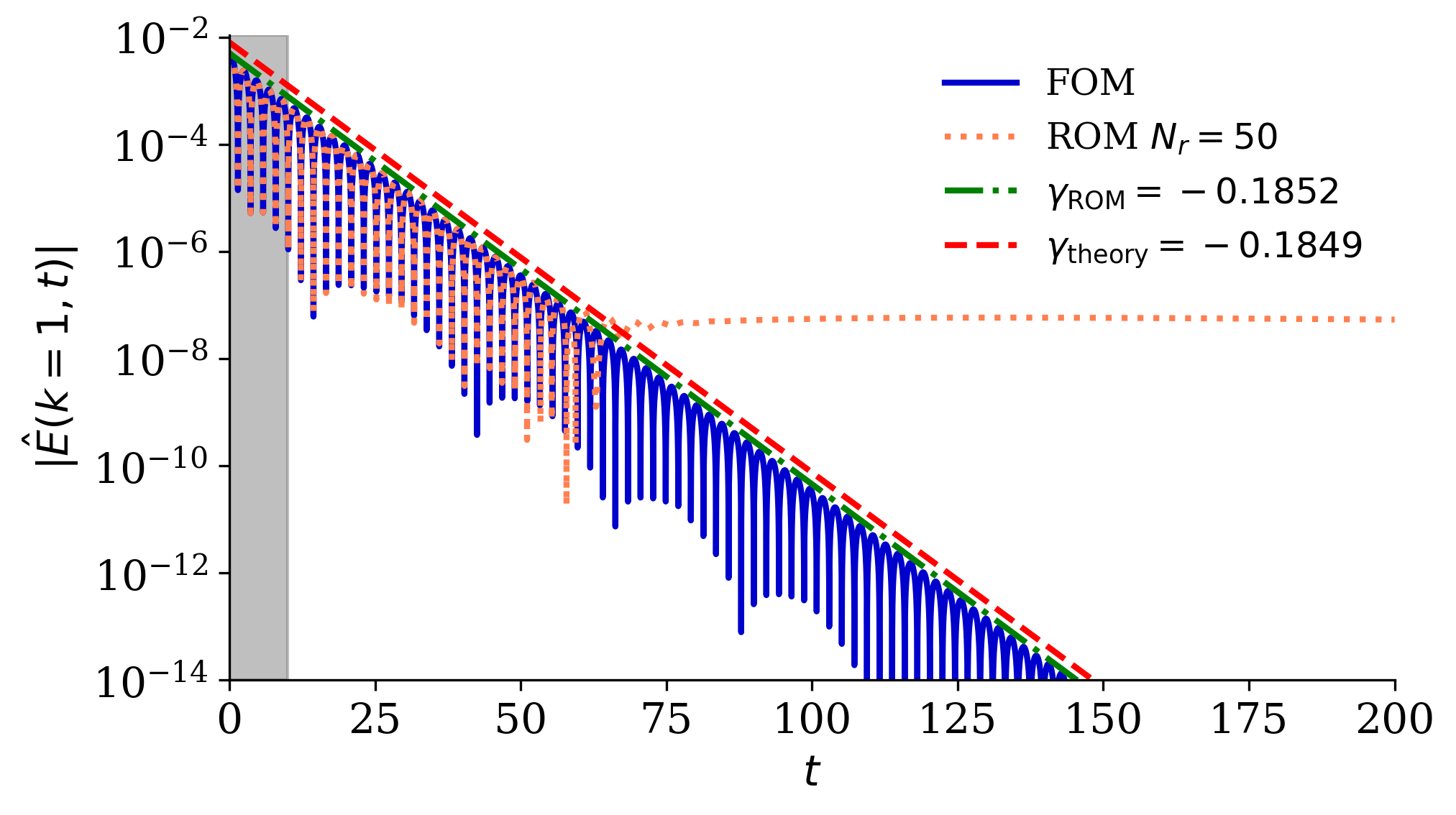}
    \end{subfigure}
    \hspace{-10pt}
    \begin{subfigure}[b]{0.5\textwidth}
        \centering 
        \caption{$\alpha_{e} = 0.5$}
        \includegraphics[width=\textwidth]{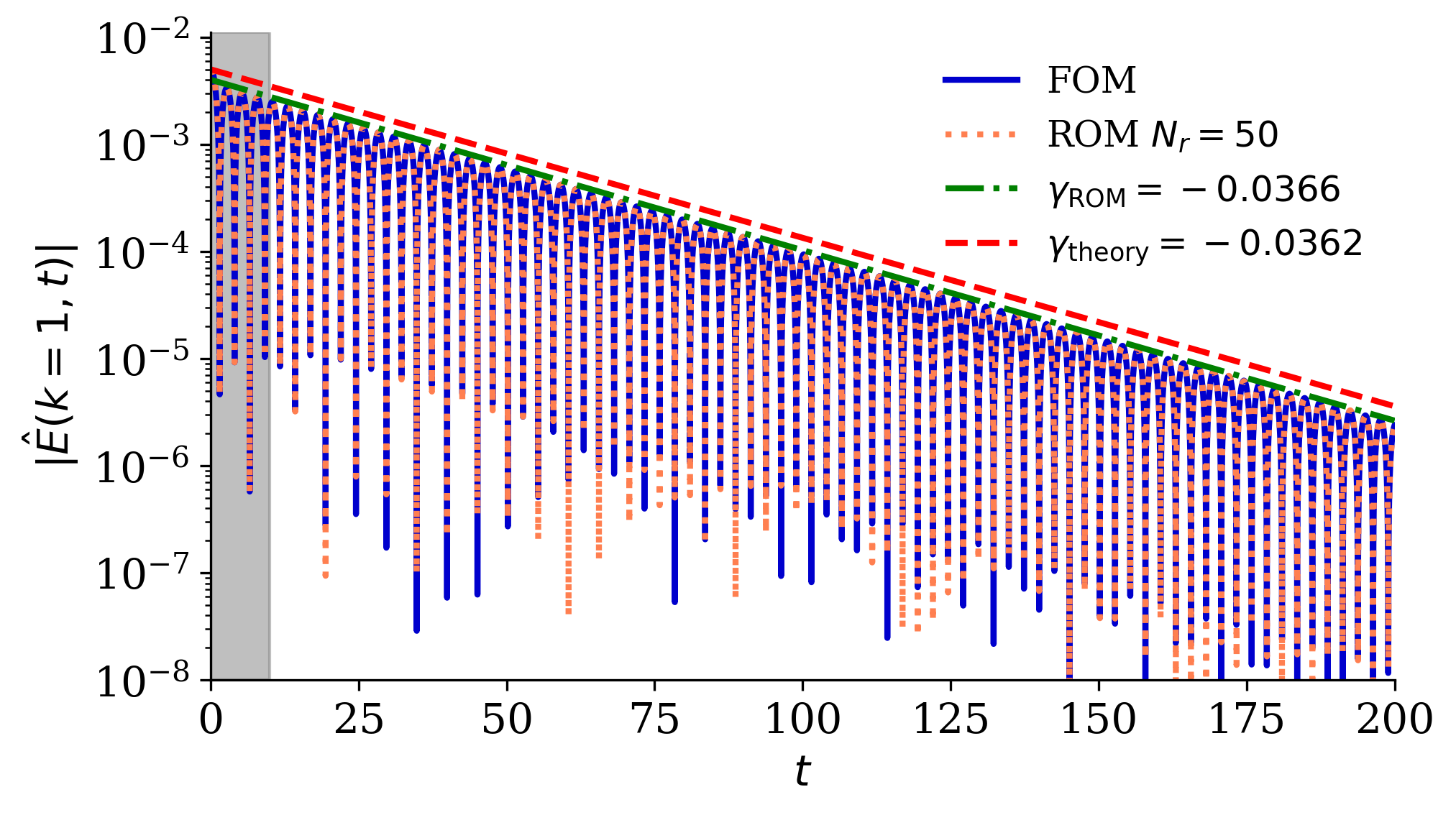}
    \end{subfigure}
    \caption{Weak Landau damping electric field first Fourier mode magnitude evolution in time for ROM  with $N_{r}=50$ for (a)~$\alpha_{e} = 0.75$ (interpolation test) and (b)~$\alpha_{e} = 0.5$ (extrapolation test). In both cases, the ROM can extrapolate parametrically and in time (training data is simulated up to $t=10$). The dashed red lines are the analytically derived damping rates from linear theory~\cite[\S 2]{gary_1993_theory} and the dashed green lines are the ROM estimated damping rates. The gray shading highlights the training time interval $t \in [0, 10]$. }
    \label{fig:weak_landau_damping_test_results}
\end{figure}

\begin{figure}
    \centering
    \begin{subfigure}[b]{0.5\textwidth}
        \centering 
        \caption{ROM with $N_{r}=50$}
        \includegraphics[width=\textwidth]{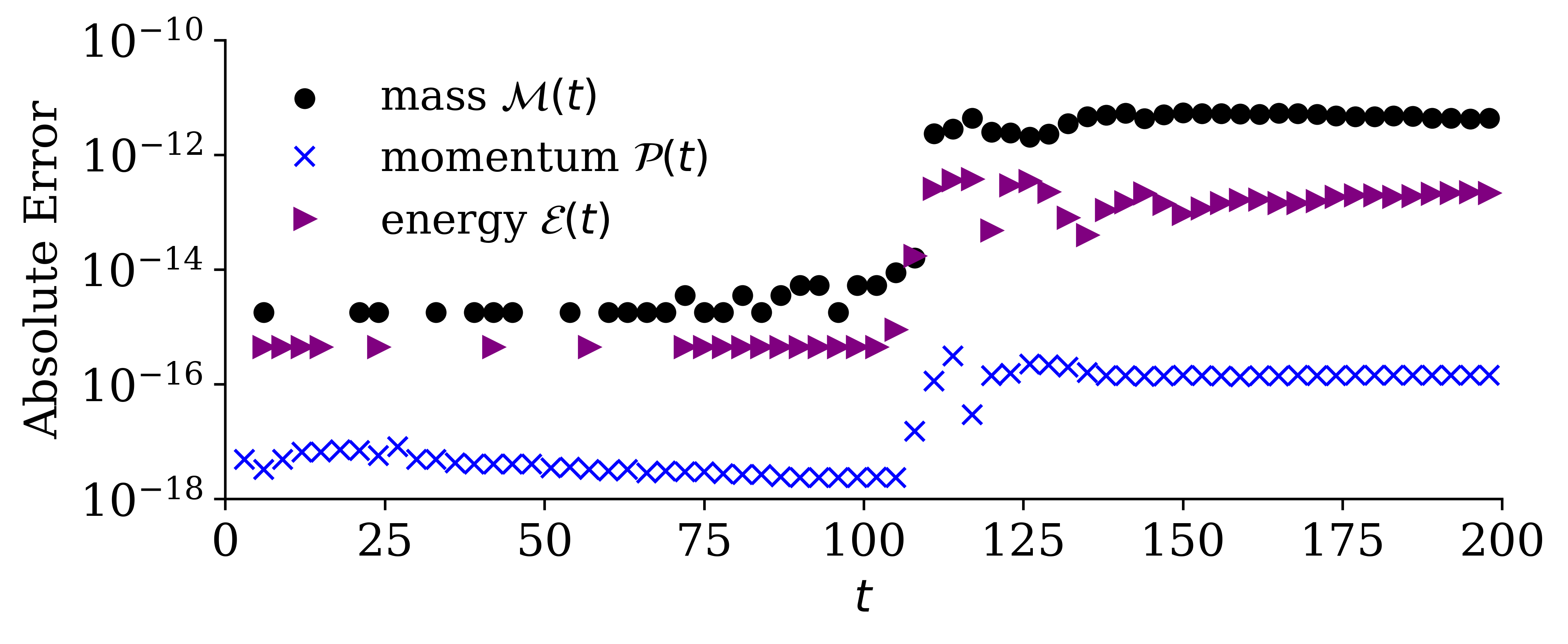}
    \end{subfigure}  
    \hspace{-10pt}
    \begin{subfigure}[b]{0.5\textwidth}
        \centering 
        \caption{FOM}
        \includegraphics[width=\textwidth]{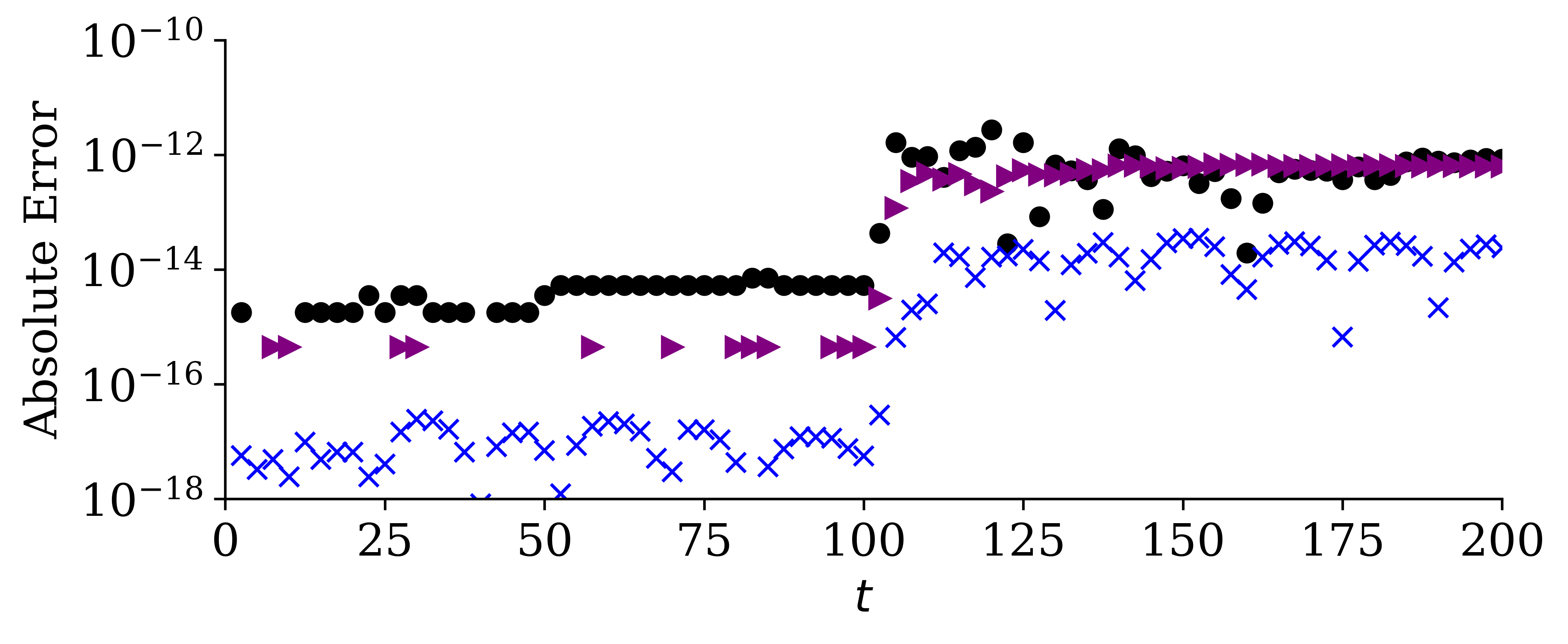}
    \end{subfigure}
    \caption{Mass, momentum, and energy conservation of weak Landau damping with $\alpha_{e}=0.5$. The (a) ROM  with $N_{r} =50$ and (b) FOM absolute errors are comparable and close to the temporal integrator nonlinear solver absolute tolerance $10^{-12}$. }
    \label{fig:weak_landau_conservation}
\end{figure}

Figure~\ref{fig:weak_landau_damping_test_results} shows the test results for $\alpha_{e} = 0.75$ (interpolation test) and $\alpha_{e} = 0.5$ (extrapolation test) with reduced dimension $N_{r} = 50$. 
In both cases, the ROM successfully predicts the electric field first Fourier mode magnitude evolution beyond the training time interval $t \in [0, 10]$.
More specifically, for $\alpha_{e} = 0.5$, the ROM can predict up $t\sim 200$, i.e. $1,900$\% beyond the training interval, and for $\alpha_{e} = 0.75$, the strongly damped case, the ROM can predict up to $t \sim 60$, i.e. $500$\% beyond the training interval. 
This observation agrees with intuition since the viable ROM simulation time is constrained by the POD basis relative error ($\sigma_{50}/\sigma_{1} \sim 10^{-6}$ for training samples, see Figure~\ref{fig:weak_landau_singular_value_decay}), such that stronger damping rate cases reach the POD error threshold faster than weaker damping rate cases.
The total number of DOFs for the FOM is $N_{v}N_{x} = 7,550$ (with $N_{K} = 7,097$) and ROM is $3N_{x} + N_{r} = 503$ (with $N_{r} = 50$), a factor $15$ in memory reduction.  
To demonstrate the ROM's effectiveness for parametric studies, we compare its estimated damping rates $\gamma_{\mathrm{ROM}}(\alpha_{e} = 0.5) = -0.0366$ and $\gamma_{\mathrm{ROM}}(\alpha_{e} = 0.75) = -0.1852$ with linear theory $\gamma_{\mathrm{theory}}(\alpha_{e} = 0.5) = -0.0362$ and $\gamma_{\mathrm{theory}}(\alpha_{e} = 0.75) = -0.1849$. The ROM damping rates are obtained via the slope of a linear least-squares fit to the peaks of $|\hat{E}(k=1, t)|$ over $t \in [0, 50]$. In contrast, simple linear interpolation/extrapolation of the theoretical damping rates from the training data (see Figure~\ref{fig:weak_landau_electric_field_amplitude}) yields less accurate estimates: $\gamma_{\mathrm{interp}}(\alpha_{e} = 0.5) = -0.0209$ and $\gamma_{\mathrm{interp}}(\alpha_{e} = 0.75) = -0.18605$. Although one would typically compute damping rates by solving the dispersion relation, this example highlights that the ROM captures the parametric dependence of the quantity of interest with significantly greater accuracy than direct interpolation. 
Moreover, Figure~\ref{fig:weak_landau_conservation} presents the ROM and FOM conservation of mass, momentum, and energy, which are comparable and close to the tolerances set by the temporal integrator nonlinear solver. The conservation errors grow at around $t=110$ after the electric field has nearly vanished, making the solution more sensitive to round-off numerical errors.

\begin{figure}
    \centering
        \begin{subfigure}[b]{0.45\textwidth}
        \centering 
        \caption{CPU runtime vs. $N_{r}$}
        \label{fig:weak_landau_CPU_time}
        \includegraphics[width=\textwidth]{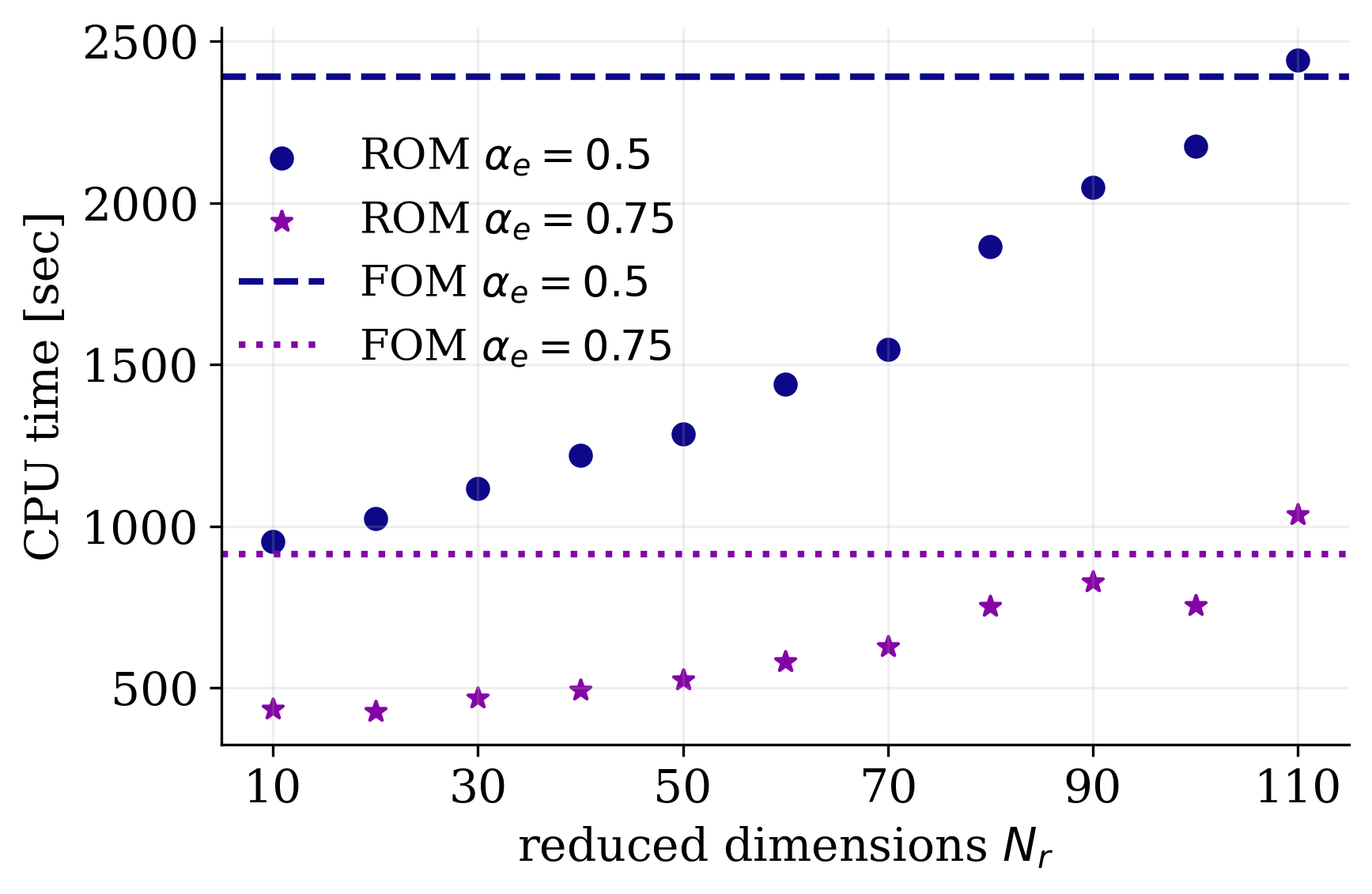}
    \end{subfigure}
    \begin{subfigure}[b]{0.45\textwidth}
        \centering 
        \caption{Density mean relative error in space and time vs. $N_{r}$}
        \label{fig:weak_landau_error}
        \includegraphics[width=\textwidth]{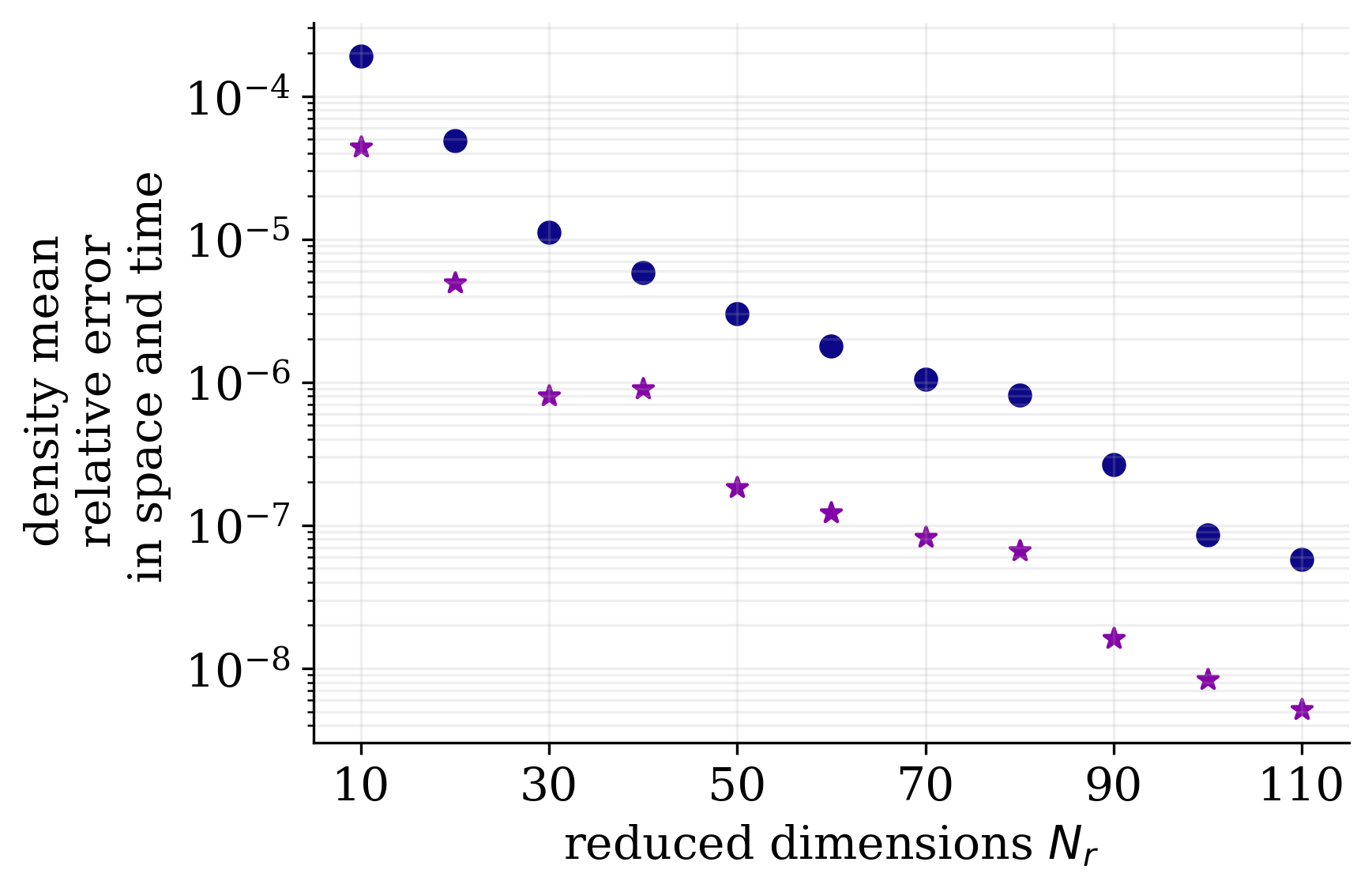}
    \end{subfigure}   
    \begin{subfigure}[b]{0.45\textwidth}
        \centering 
        \caption{Density mean relative error in space for $\alpha_{e} = 0.5$}
        \label{fig:weak_landau_error_05}
        \includegraphics[width=\textwidth]{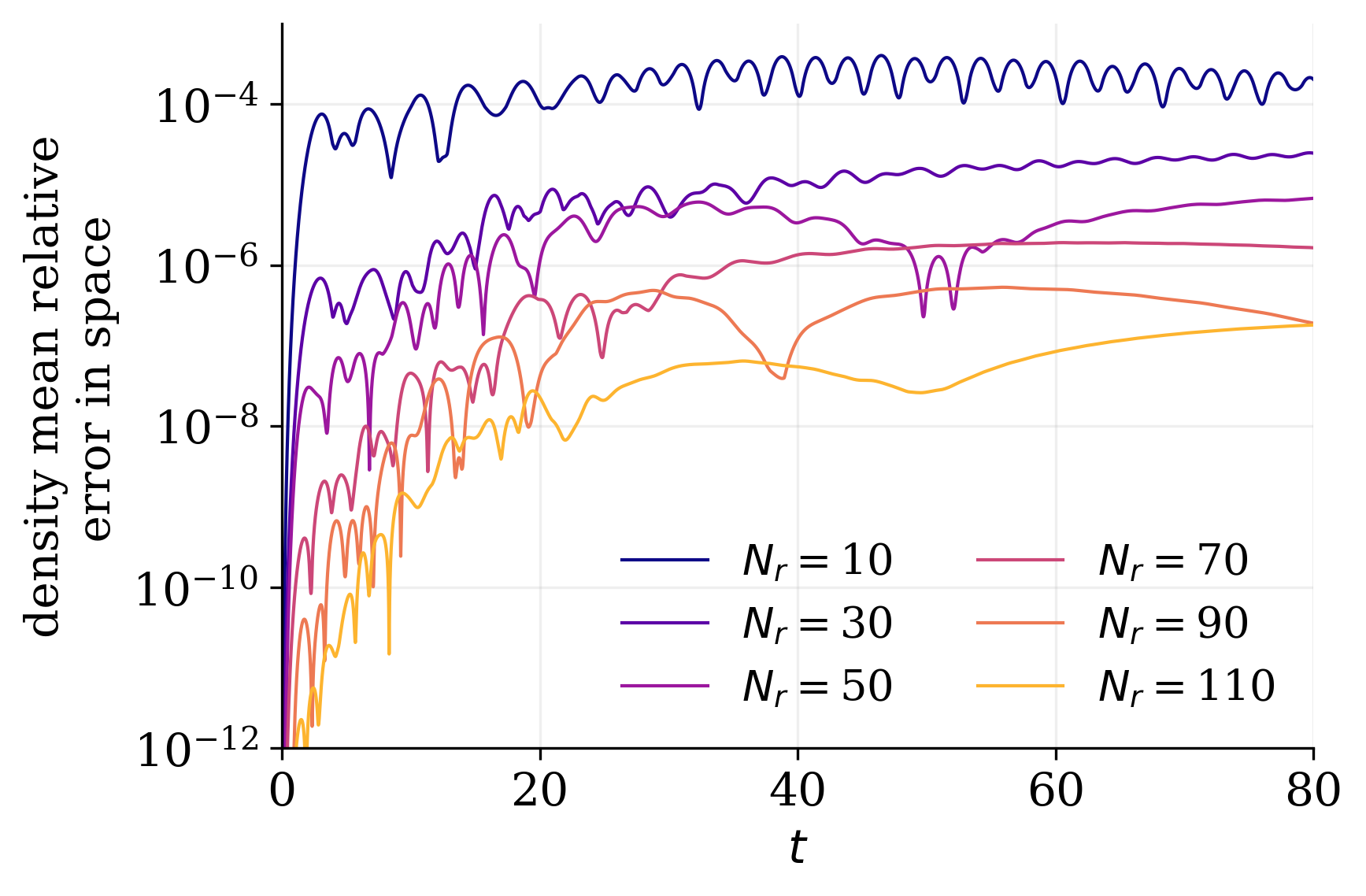}
    \end{subfigure}  
    \begin{subfigure}[b]{0.45\textwidth}
        \centering 
        \caption{Density mean relative error in space for $\alpha_{e} = 0.75$}
        \label{fig:weak_landau_error_075}
        \includegraphics[width=\textwidth]{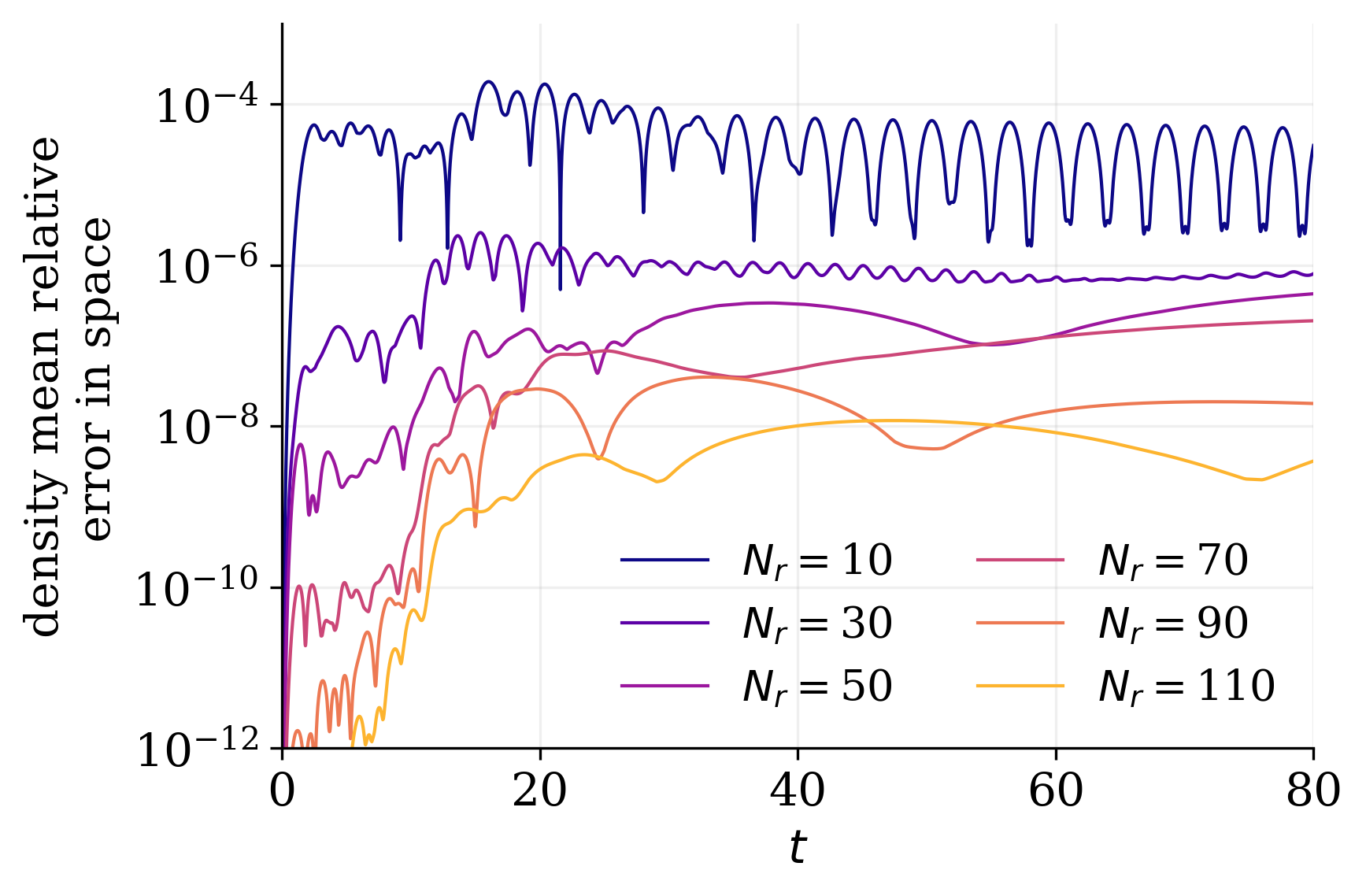}
    \end{subfigure}  
    \caption{Weak Landau damping ROM (a) CPU runtime and (b) density mean relative error in space and time for out-of-training extrapolated sample $\alpha_{e} = 0.5$ and interpolated sample $\alpha_{e} = 0.75$, and density mean relative error in space (varying the time) for (c) $\alpha_{e} = 0.5$ and (d) $\alpha_{e} = 0.75$. As expected, increasing the ROM dimensions increases the CPU runtime and improves the models' accuracy. The ROM eventually becomes more computationally expensive than the FOM (around $N_{r}\sim 110$) since the FOM, despite its high dimensionality, employs sparse operators, whereas the ROM, although lower-dimensional, is based on dense operators. Moreover, the relative error remains stagnant in time, which is important for long-term simulation capabilities.}
    \label{fig:weak_landau_performance}
\end{figure}

We analyze the ROM performance for a range of reduced dimensions $N_{r}$ for $\alpha_{e} = 0.5$ and $\alpha_{e}=0.75$. As expected, Figure~\ref{fig:weak_landau_CPU_time} shows that increasing the reduced dimension $N_{r}$ increases the ROM CPU runtime. 
The runtime differences between the two cases stem from the nonlinear temporal solver, which required more iterations for $\alpha_{e} = 0.5$ in each time step compared to $\alpha_{e} = 0.75$, for both the ROM and FOM simulations. 
The weak Landau damping test case is relatively simple and involves a small number of FOM DOFs, resulting in ROM with $N_{r}=50$ and FOM CPU runtimes differing by only a factor of about $2$. 
It is important to note that as $N_{r}$ increases, the ROM eventually surpasses the FOM in computational cost (at $N_{r} \sim 110$), as the FOM is high-dimensional and involves sparse operators, whereas the ROM is low-dimensional and involves dense operators.
Figure~\ref{fig:weak_landau_error} shows the electron density $n_{e}(x, t) \coloneqq \alpha_{e} C_{e, 0}(x, t)$ mean relative error in space $x \in [0, 2\pi]$ and time $t\in [0, 80]$ of the ROM:
\begin{equation}\label{density_mean_relative_error}
    \frac{1}{N_{t} N_{x}}\sum_{j=0}^{N_{t}} \sum_{i=0}^{N_{x}}\frac{|n^{\mathrm{ROM}}_{e}(x_{i}, t_{j}) - n^{\mathrm{FOM}}_{e}(x_{i}, t_{j})|}{|n^{\mathrm{FOM}}_{e}(x_{i}, t_{j})|}.
\end{equation}
The results show that the mean relative error in space and time reduces exponentially as a function of the reduced dimensions $N_{r}$. Additionally, Figures~\ref{fig:weak_landau_error_05}--\ref{fig:weak_landau_error_075} show the density mean relative error in space (varying in time) for $\alpha_{e}=0.5$ and $\alpha_{e} = 0.75$. For both parametric cases, the relative error remains stagnant in time, which is important for long-term simulation capabilities. 

\subsection{Two-Stream Instability} \label{sec:two-stream-instability-results}
\begin{figure}
    \centering
        \begin{subfigure}[b]{0.325\textwidth}
        \centering 
        \caption{Two-stream initial condition}
        \includegraphics[width=\textwidth]{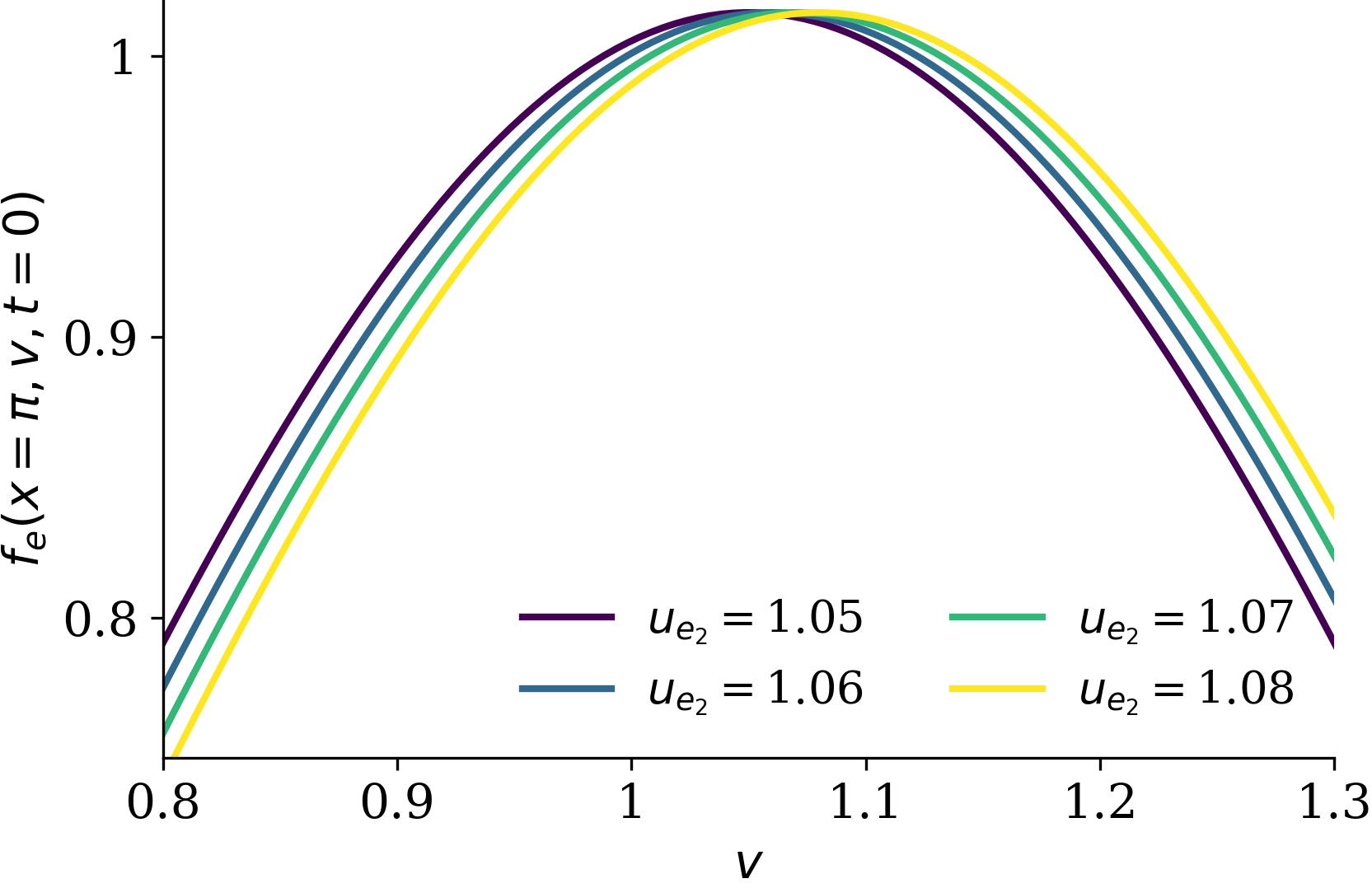}
    \end{subfigure}
    \hspace{-5pt}
    \begin{subfigure}[b]{0.325\textwidth}
        \centering 
        \caption{Two-stream electric field first Fourier mode magnitude}
        \includegraphics[width=\textwidth]{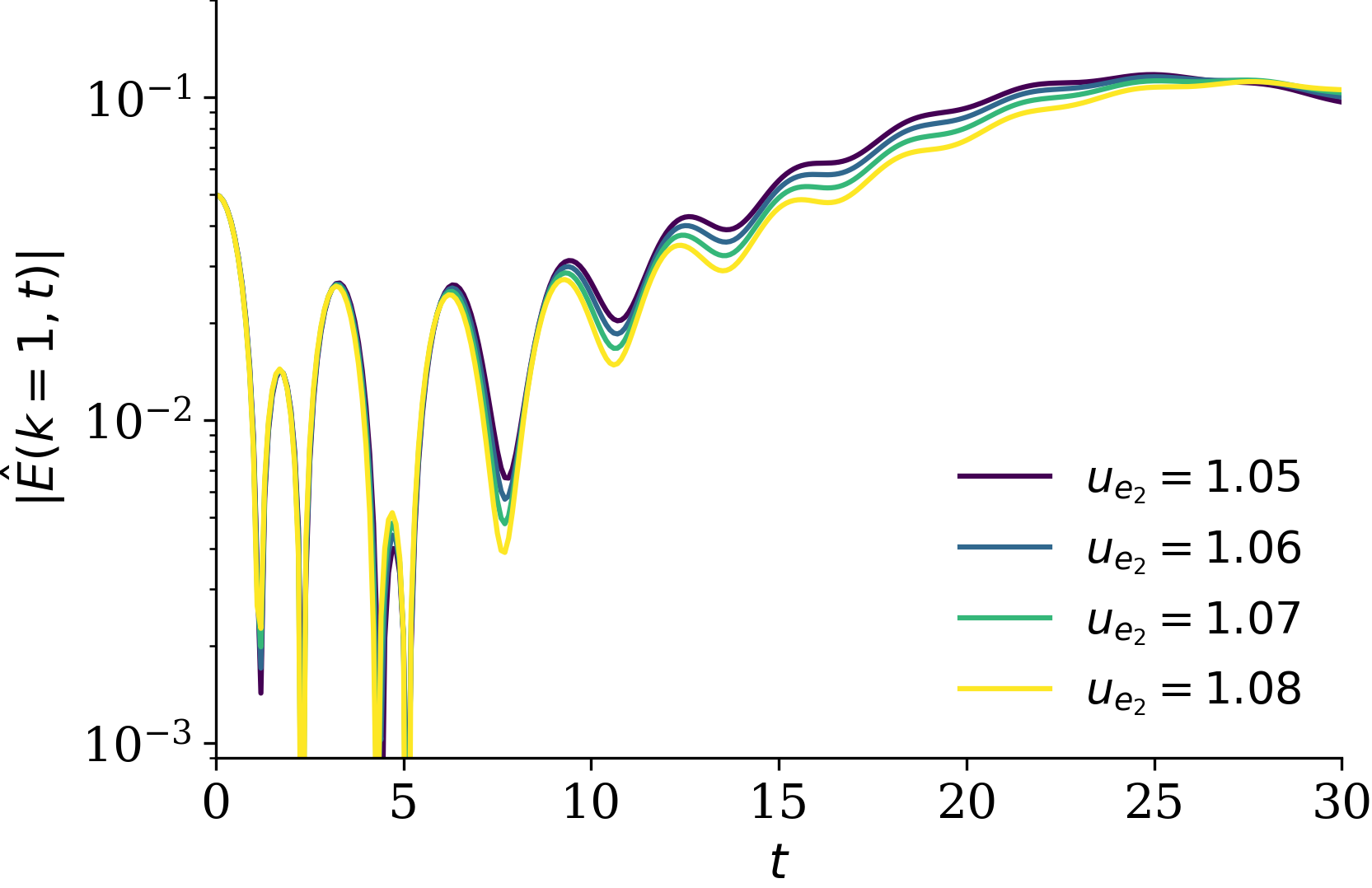}
    \end{subfigure}
    \hspace{-5pt}
    \begin{subfigure}[b]{0.325\textwidth}
        \centering 
        \caption{Two-stream singular value decay}
        \includegraphics[width=\textwidth]{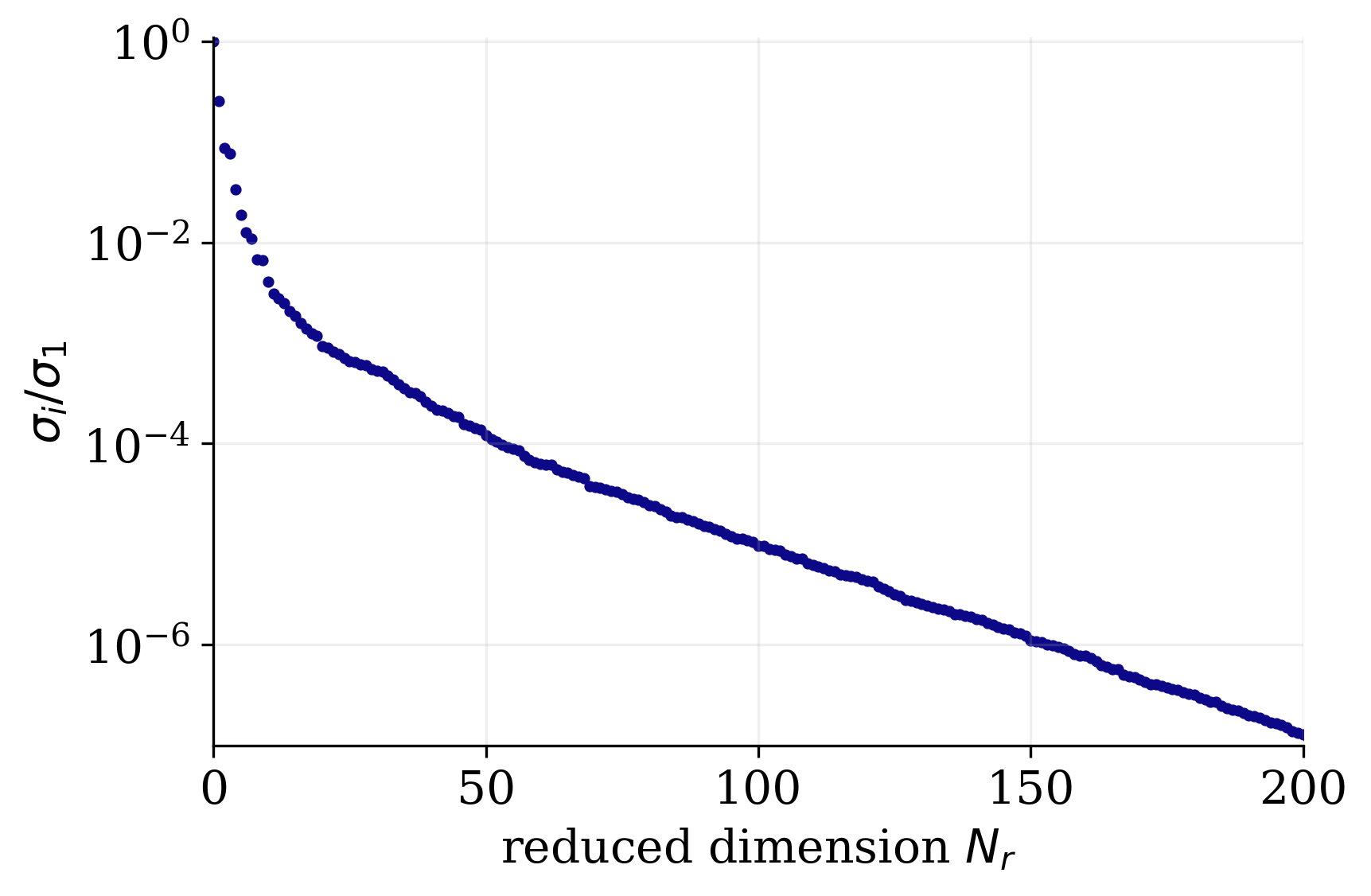}
    \end{subfigure}
    \caption{Same as Figure~\ref{fig:weak_landau_damping_setup} for the two-stream instability. The singular value decay shows that a significant reduction is viable since $\sigma_{150} = 10^{-6} \sigma_{1}$ and the FOM kinetic dimensions is $N_{K} = (N_{v} -3) N_{x} = 87,097$.}
    \label{fig:two_stream_setup}
\end{figure}

\begin{figure}
    \centering
     \includegraphics[width=\textwidth]{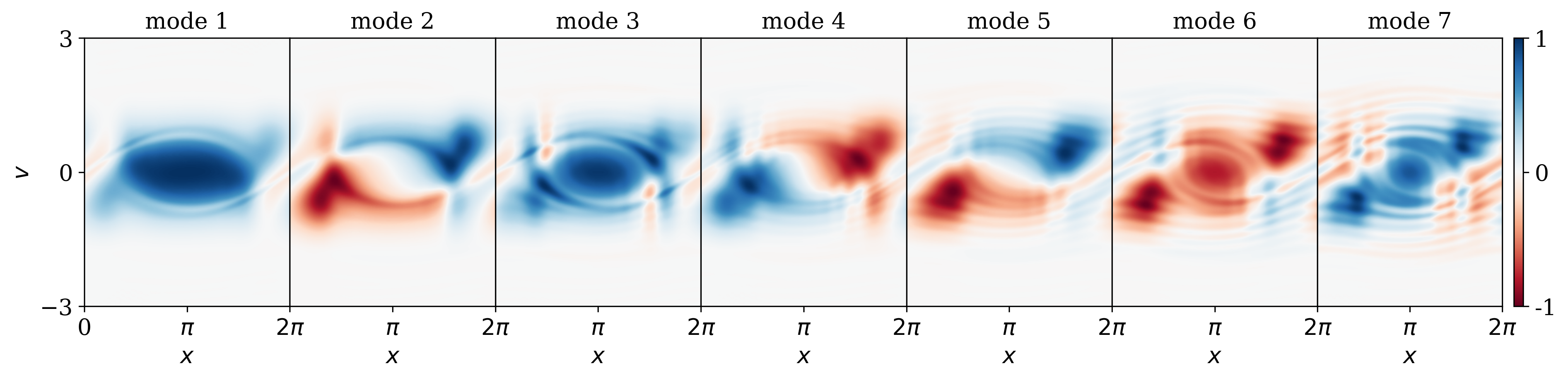}
    \caption{Two stream instability first seven POD modes in phase space (normalized to respective maximum value). The higher modes capture small-scale spatial and velocity structures around the electron distribution function phase space vortex.}
    \label{fig:two_stream_pod_modes}
\end{figure}

We simulate the two-stream instability with a relatively high amplitude perturbation $\epsilon = 0.1$ as a strongly nonlinear benchmark problem. We simulate the system for $t \in [0, 30]$. In this parametric study, we vary the velocity of each electron beam such that $u_{e_{2}} = -u_{e_{1}} = \{1.05, 1.06, 1.07, 1.08\}$. The two electron beams are evolved as distinct species, while the ions are treated as a static neutralizing background. The initial conditions to generate the training data, the electric field first Fourier mode magnitude, and the decay of the singular value are shown in Figure~\ref{fig:two_stream_setup}. The singular value decay indicates that $\sigma_{150} = 10^{-6} \sigma_{1}$, which implies that there is a low-dimensional linear subspace representation of the kinetic state as the kinetic dimension of the FOM is $N_{K} = (N_{v}-3)N_{x} = 87,097$. The first seven normalized POD modes are shown in Figure~\ref{fig:two_stream_pod_modes}. The higher modes capture small-scale structures in phase space around the vortex that forms between the two electron beams.  

\begin{figure}
    \centering
        \begin{subfigure}[b]{0.2\textwidth}
        \centering 
        \caption{FOM}
        \includegraphics[width=\textwidth]{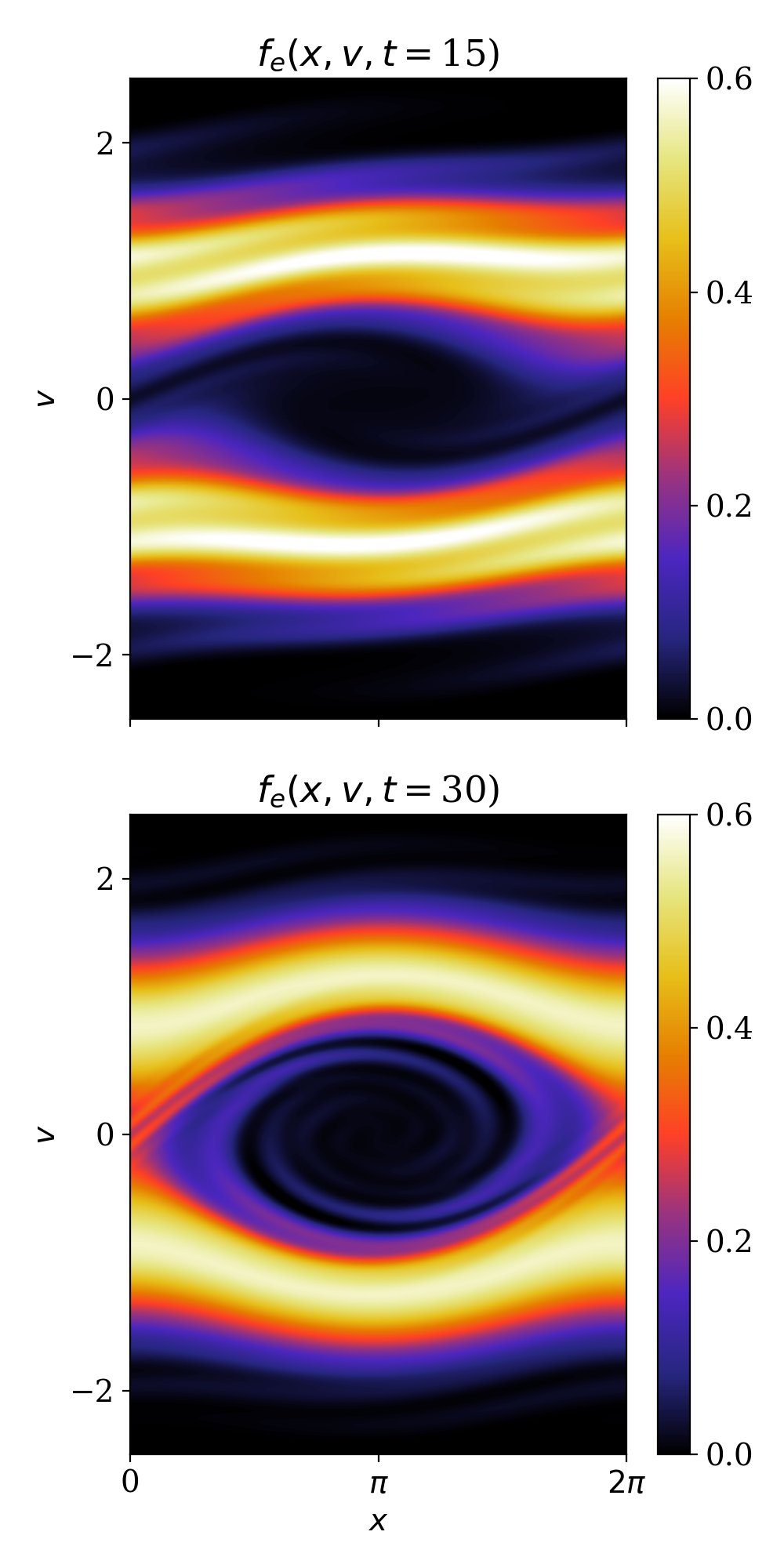}
    \end{subfigure}
    \hspace{-8pt}
    \begin{subfigure}[b]{0.2\textwidth}
        \centering 
        \caption{ROM $N_{r}=150$}
        \includegraphics[width=\textwidth]{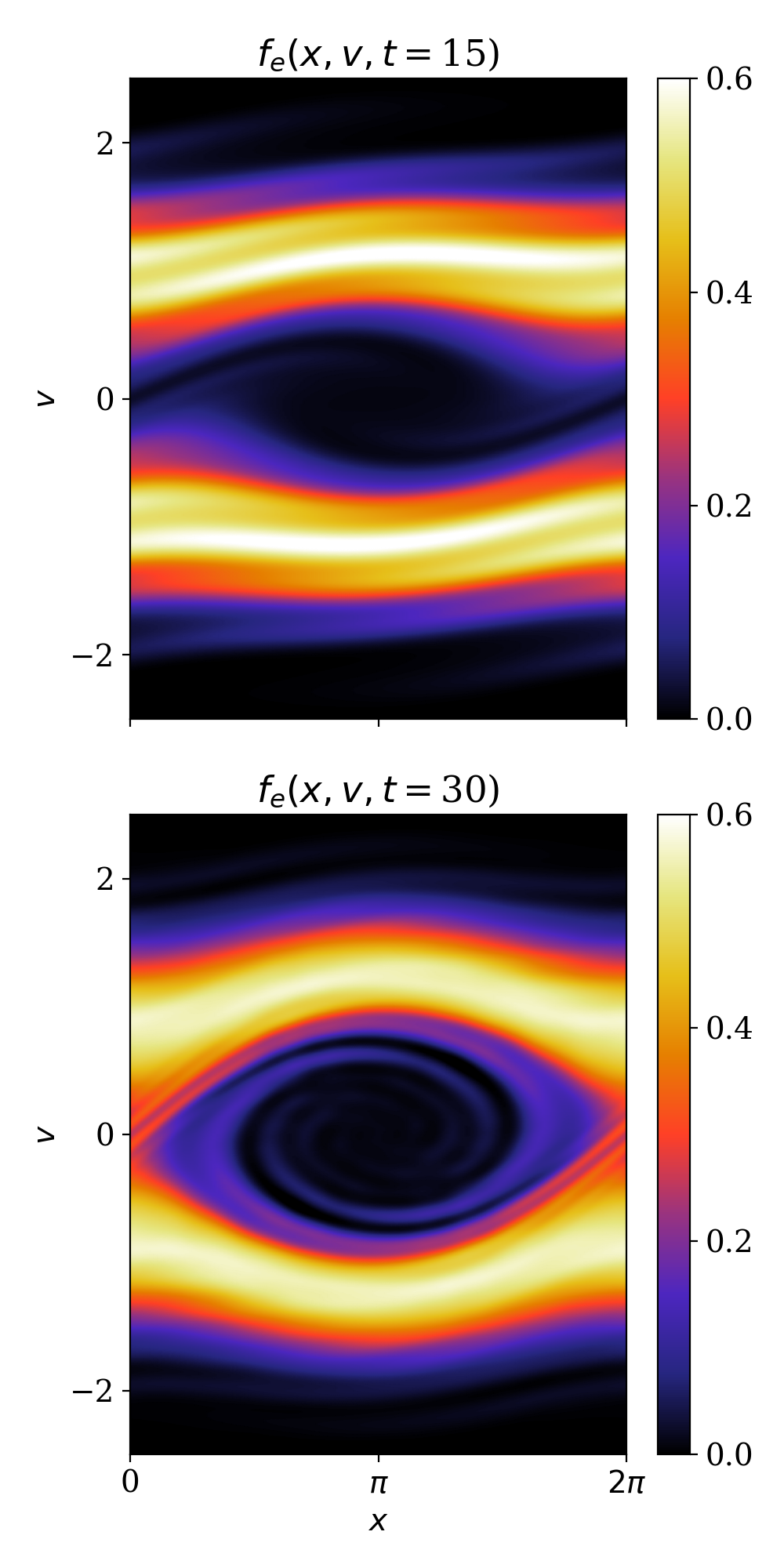}
    \end{subfigure}
    \hspace{-8pt}
    \begin{subfigure}[b]{0.2\textwidth}
        \centering 
        \caption{Abs error $N_{r}=150$}
        \includegraphics[width=\textwidth]{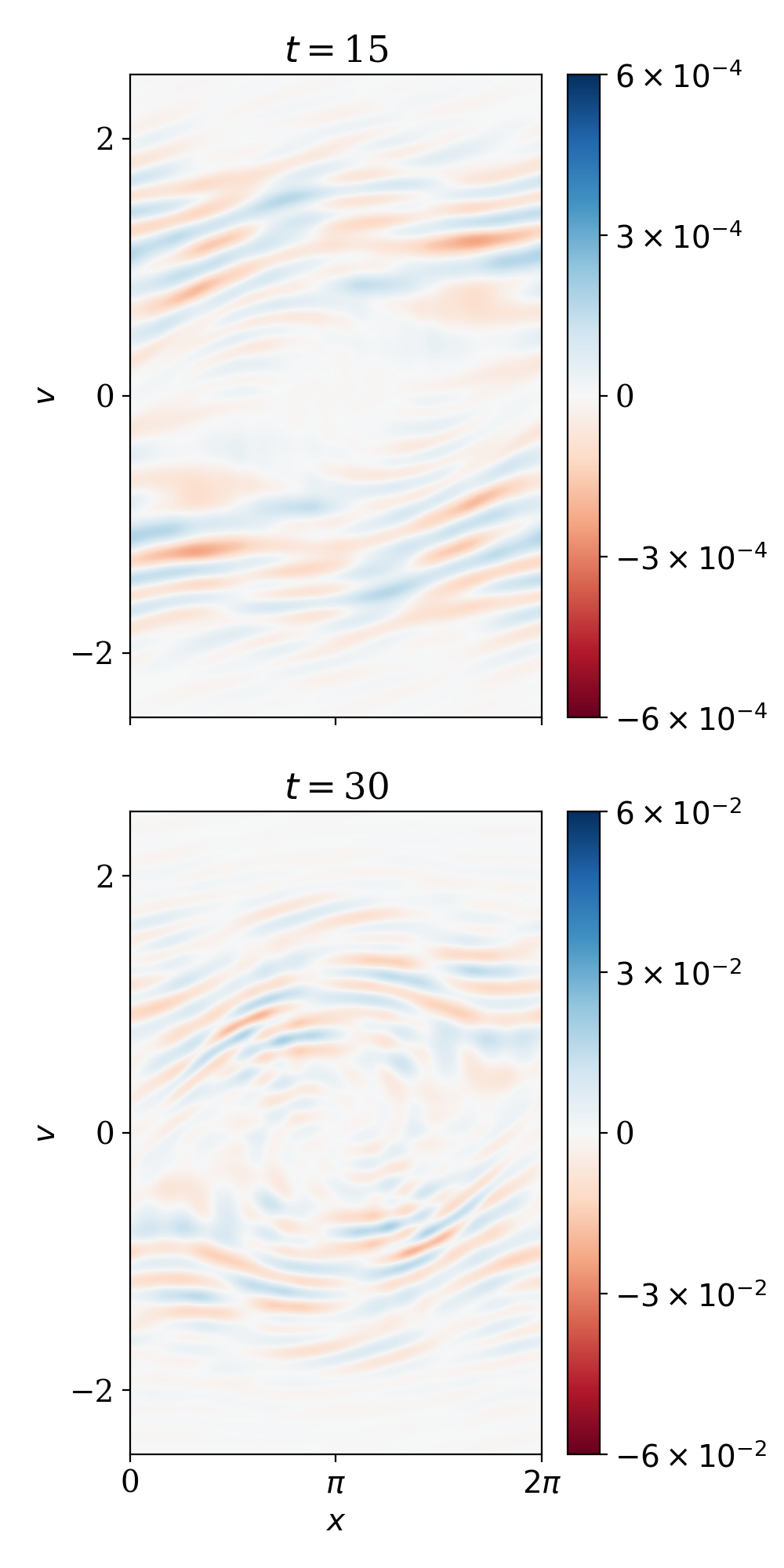}
    \end{subfigure}
    \hspace{-8pt}
    \begin{subfigure}[b]{0.2\textwidth}
        \centering 
        \caption{ROM $N_{r}=120$}
        \includegraphics[width=\textwidth]{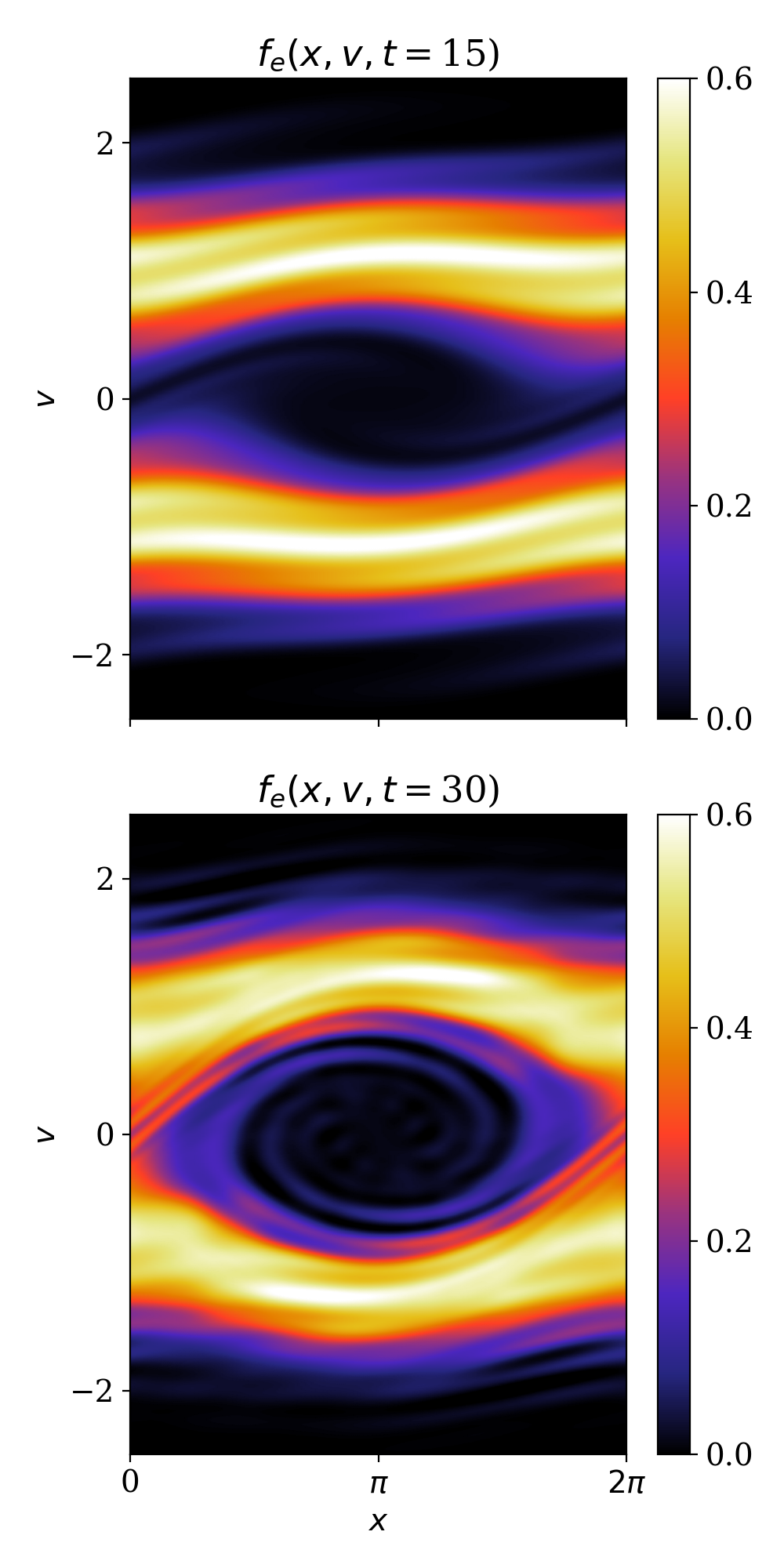}
    \end{subfigure}
    \hspace{-8pt}
    \begin{subfigure}[b]{0.2\textwidth}
        \centering 
        \caption{Abs error $N_{r}=120$}
        \includegraphics[width=\textwidth]{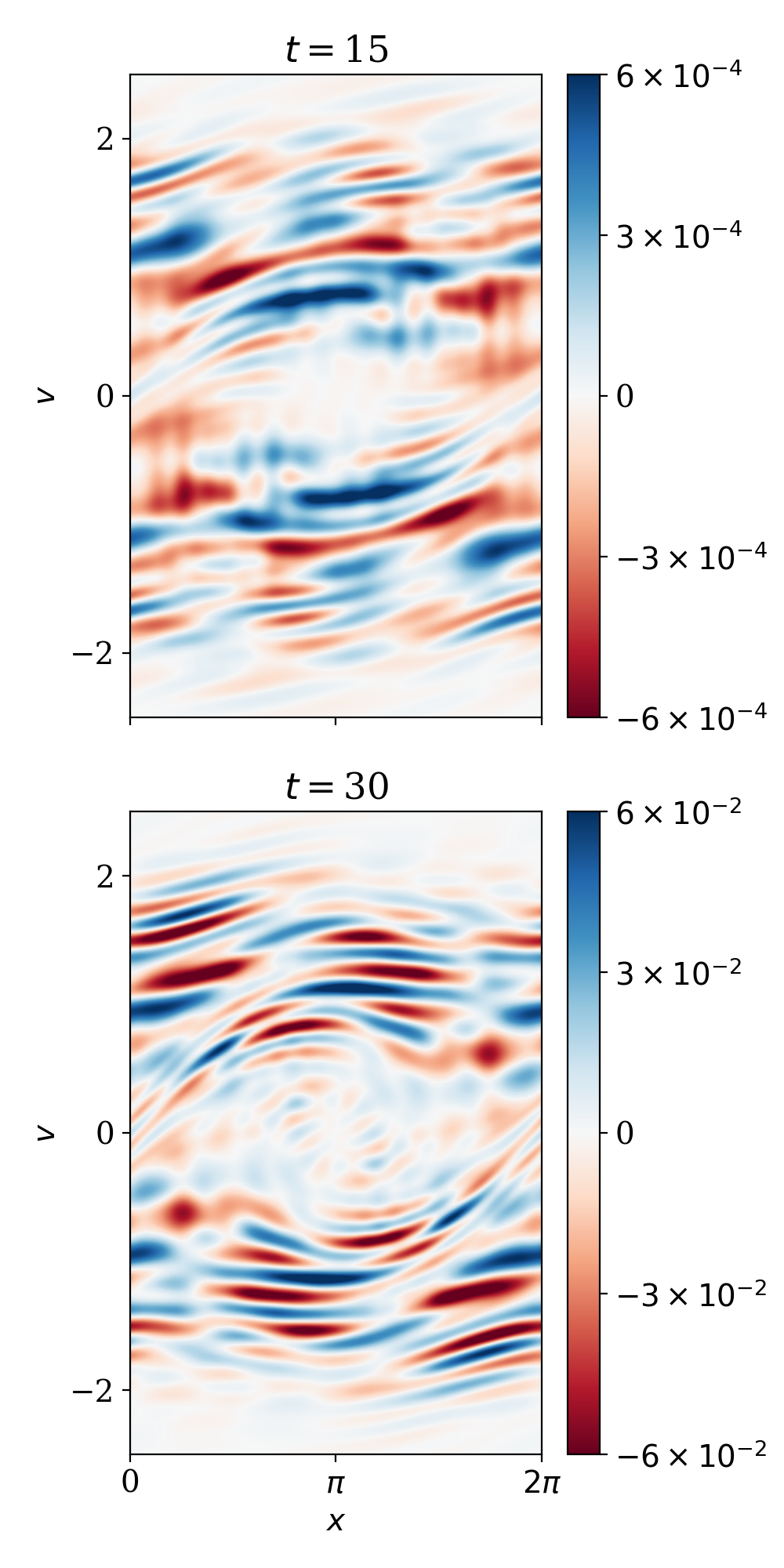}
    \end{subfigure}
    \caption{Two stream instability comparison of the electron distribution function in phase space for an interpolated value of $u_{e_{2}}=1.065$ of (a) FOM, (b) ROM with $N_{r} = 150$, and (d) ROM with $N_{r} = 120$, along with absolute error of ROM with (c) $N_{r} = 150$ and (e) $N_{r} = 120$. The results show that the ROM requires $N_{r} \approx 150 >120$ to properly predict the electron distribution function up to $t=30$. }
    \label{fig:two_stream_distribution_function_comparison}
\end{figure}

\begin{figure}
    \centering
    \begin{subfigure}[b]{0.45\textwidth}
        \centering 
        \caption{Electric field first Fourier mode magnitude comparison}
        \includegraphics[width=\textwidth]{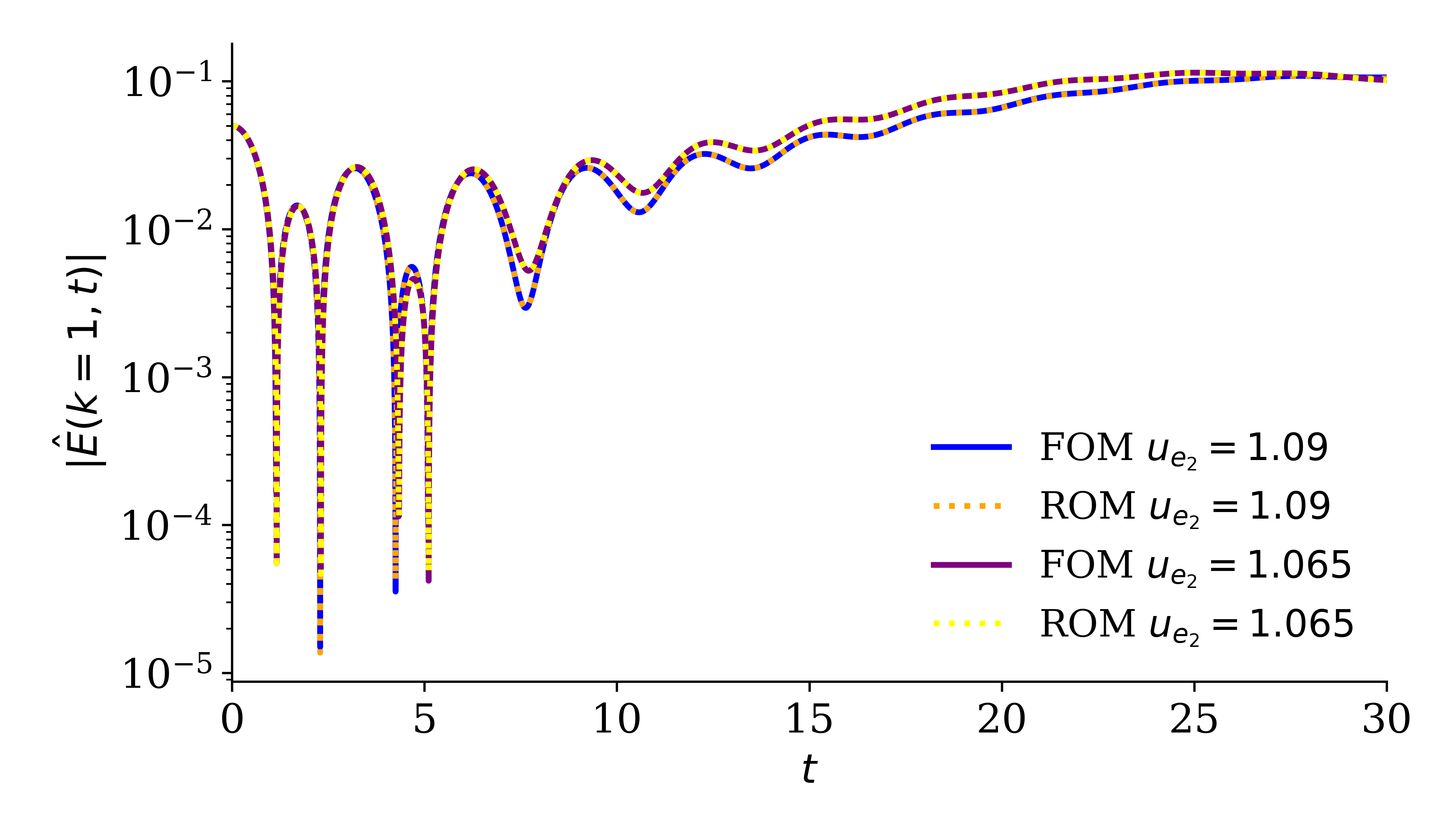}
    \end{subfigure} 
    \begin{subfigure}[b]{0.45\textwidth}
        \centering 
        \caption{Electric field first Fourier mode magnitude relative error}
        \includegraphics[width=\textwidth]{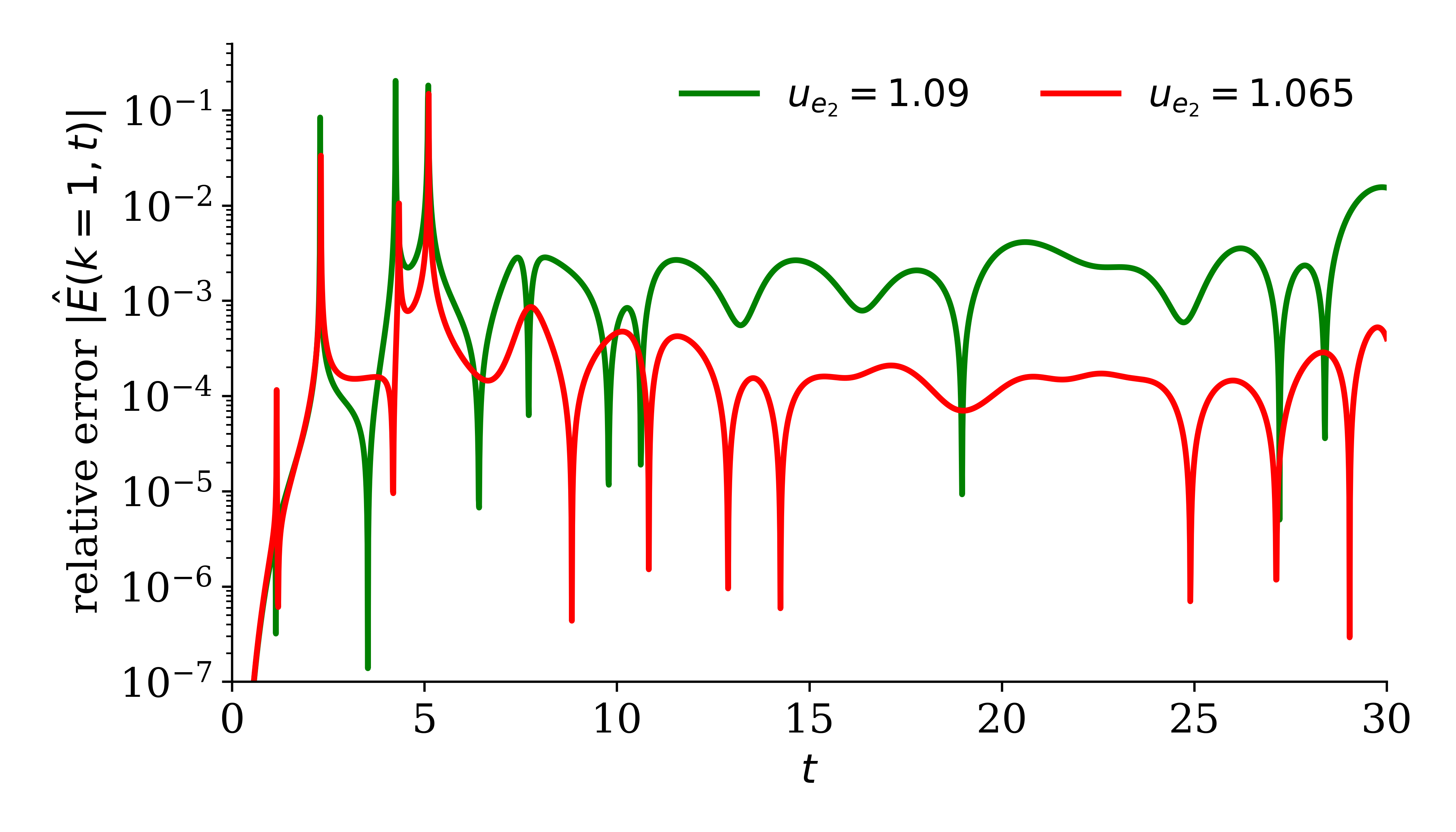}
    \end{subfigure} 
    \caption{Two stream instability comparison of the FOM and ROM with $N_{r} =150$ electric field first Fourier mode magnitude evolution evolution in time for samples $u_{e_{2}} = 1.065$ (interpolation test) and $u_{e_{2}} = 1.09$ (extrapolation test) is shown in subfigure~(a). Subfigure~(b) illustrates the respective relative error. In both cases, the relative error is stable throughout the simulation duration.}
    \label{fig:two_stream_growth_rate_comparison}
\end{figure}

\begin{figure}
    \centering
    \begin{subfigure}[b]{0.45\textwidth}
        \centering 
        \caption{ROM with $N_{r}=150$}
        \includegraphics[width=\textwidth]{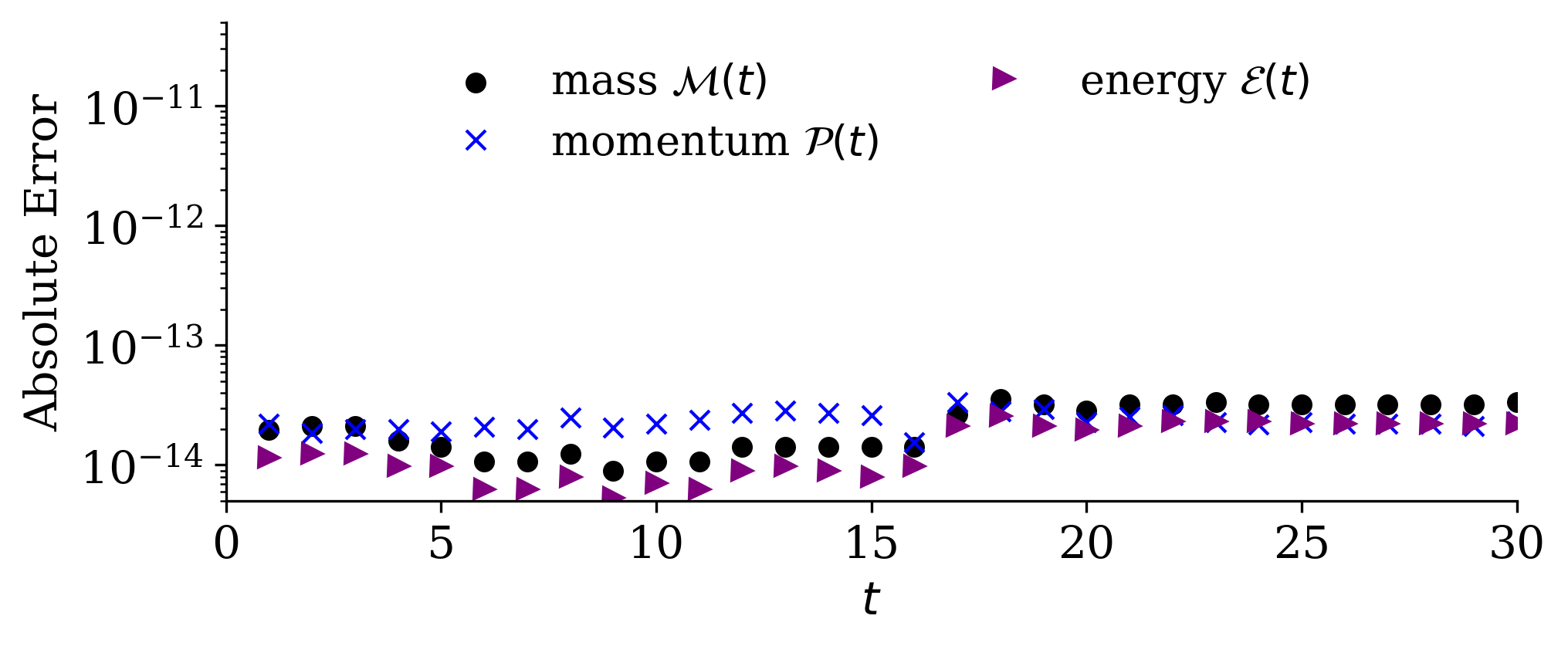}
    \end{subfigure} 
    \hspace{-10pt} 
    \begin{subfigure}[b]{0.45\textwidth}
        \centering 
        \caption{FOM}
        \includegraphics[width=\textwidth]{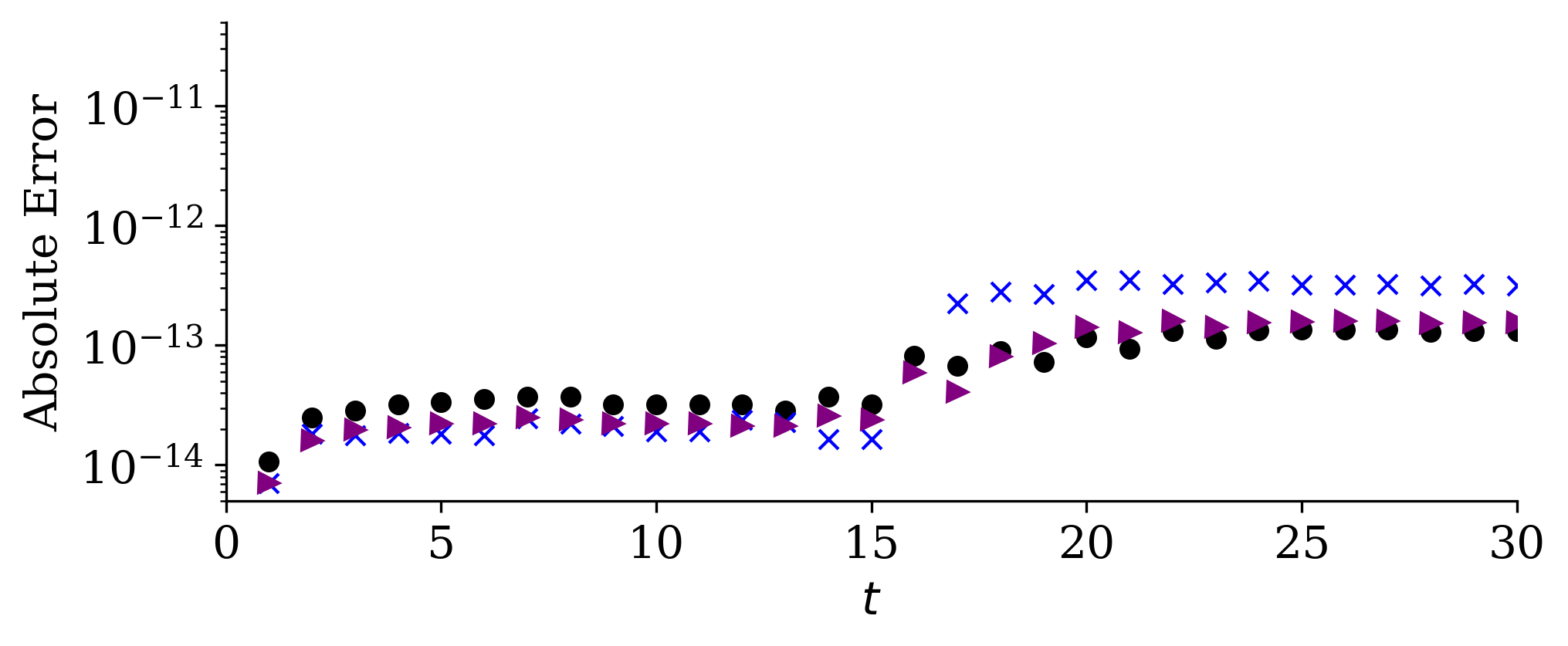}
    \end{subfigure}
    \caption{Mass, momentum, and energy conservation of the two-stream instability test case with $u_{e_{2}} = 1.065$. The (a) ROM with $N_r = 150$ and (b) FOM show comparable absolute conservation errors, both close to the temporal integrator nonlinear solver's absolute tolerance of $10^{-12}$, validating the analytic results in section~\ref{sec:conservation_properties}.}
    \label{fig:two_stream_conservation}
\end{figure}

The electron distribution function in phase space for an interpolation sample $u_{e_{2}} = 1.065$ is shown at times $t = 15$ and $t = 30$ in Figure~\ref{fig:two_stream_distribution_function_comparison} for the FOM and the ROM with $N_{r} = 120$ and $N_{r} = 150$ along with absolute errors. The results indicate that $N_{r} = 150$ accurately predicts the electron distribution function, while $N_{r} = 120$ is insufficient for predictions up to $t = 30$, where the absolute error differs by a factor of $4$ between the ROM with $N_{r} = 150$ and $N_{r} = 120$. 
In contrast to the weak Landau damping results, the two-stream instability ROM cannot accurately predict beyond the training interval in time and can only provide parametric predictions.
This limitation stems from the nonlinear dynamics, which prevents the development of a global reduced basis valid over long times. We compare the electric field first Fourier mode magnitude for an interpolation sample $u_{e_{2}} = 1.065$ and an extrapolation sample $u_{e_{2}} = 1.09$  of the FOM and ROM with $N_{r}=150$ in Figure~\ref{fig:two_stream_growth_rate_comparison}. The results show that the relative error stagnates. 
The memory footprint of the simulation reduces significantly since the total number of DOFs of the FOM is $2N_{v}N_{x} = 175,700$ and the ROM uses $N_{r}=150$, so that the total number of DOFs is $2(3N_{x} +N_{r}) = 1,806$, a $97$ factor reduction.
Lastly, Figure~\ref{fig:two_stream_conservation} confirms the conservation laws derived analytically in section~\ref{sec:conservation_properties} for both the ROM and FOM. The conservation errors grow at around $t=15$ when the simulation enters the strongly nonlinear stage of the dynamics, where mode coupling and fine-scale structures can amplify numerical errors. 

\begin{figure}
    \centering
        \begin{subfigure}[b]{0.45\textwidth}
        \centering 
        \caption{CPU time vs. $N_{r}$}
        \includegraphics[width=\textwidth]{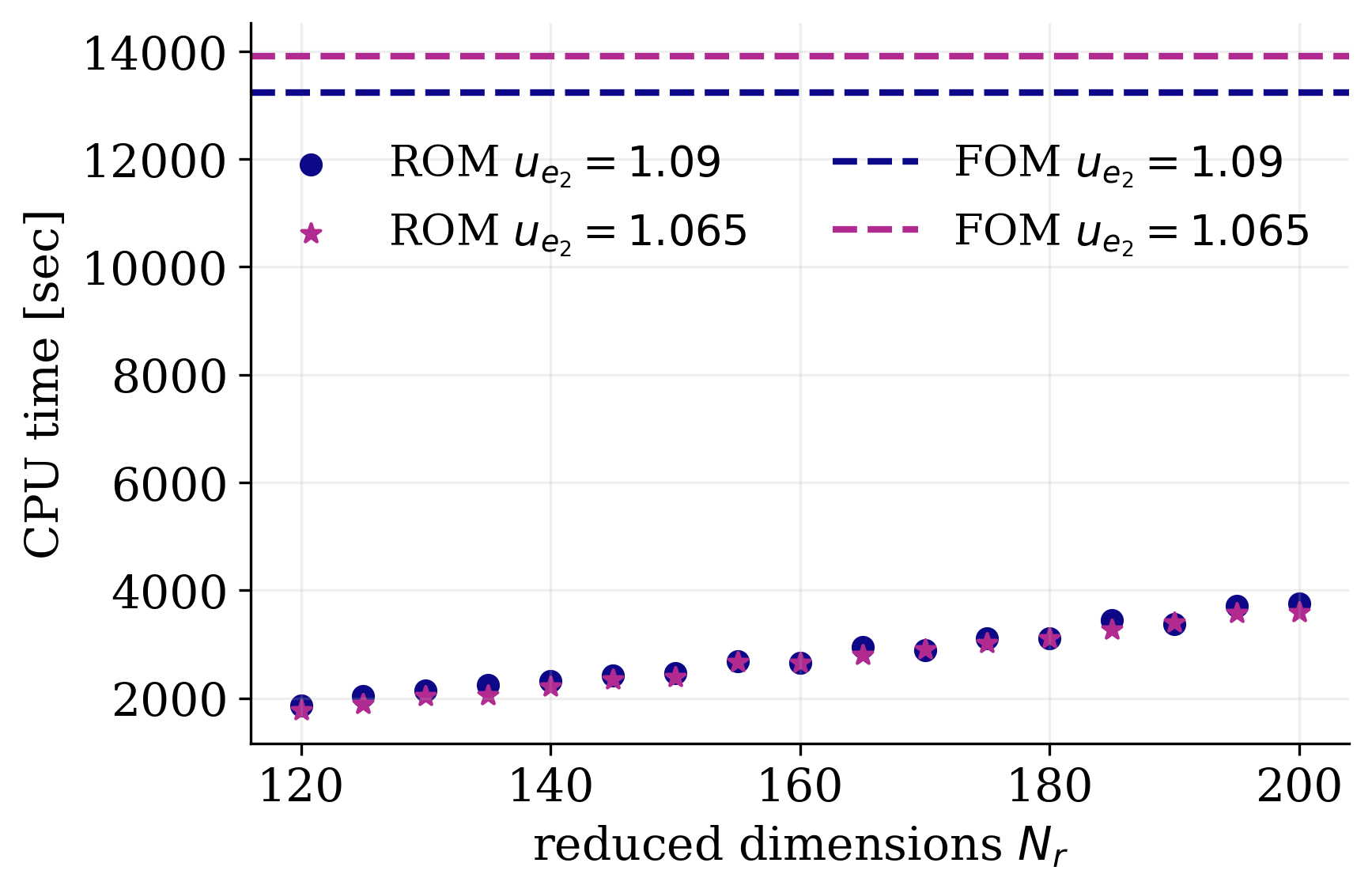}
    \end{subfigure}
    \begin{subfigure}[b]{0.45\textwidth}
        \centering 
        \caption{Density mean relative error in space and time vs. $N_{r}$}
        \includegraphics[width=\textwidth]{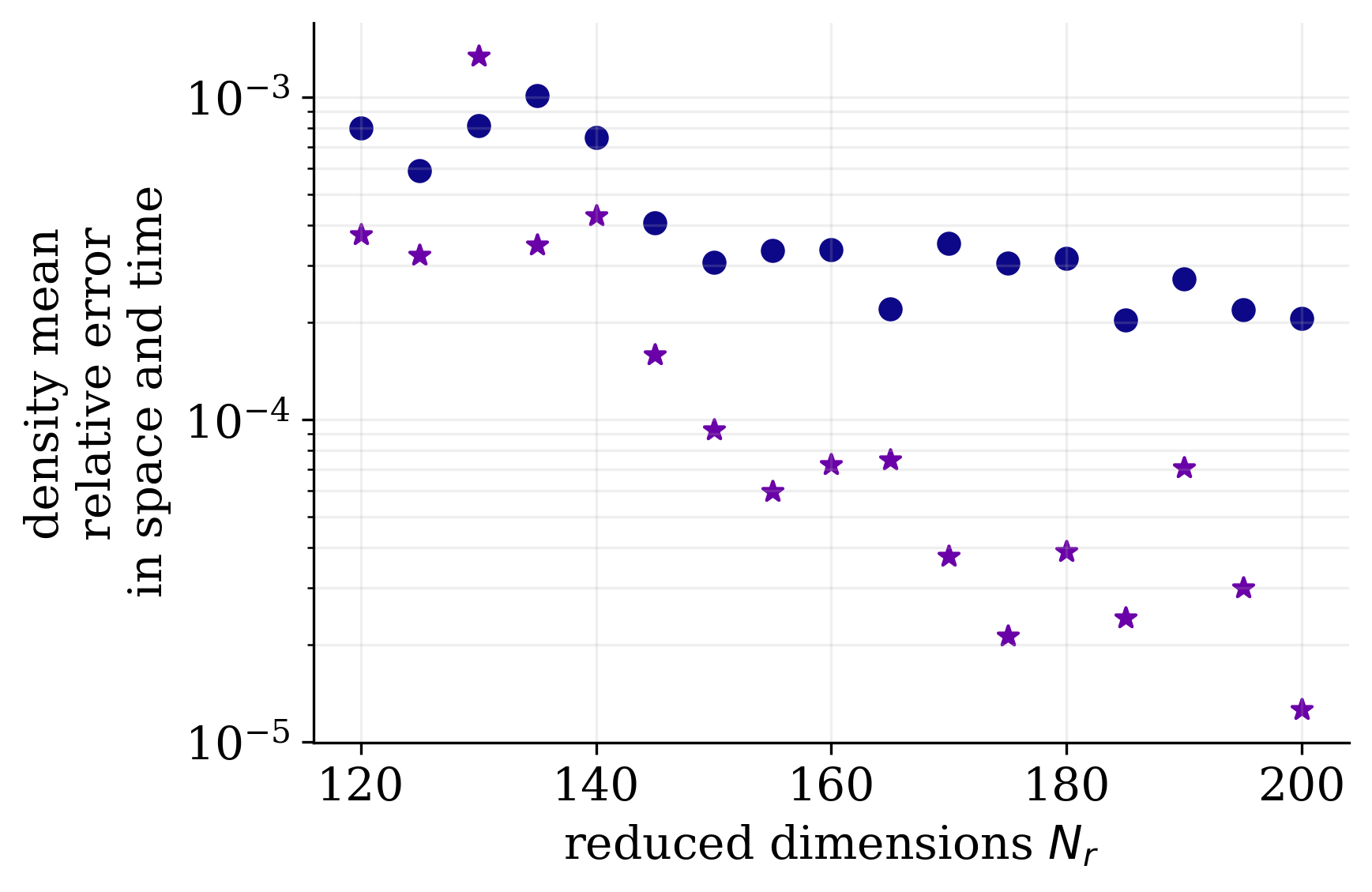}
    \end{subfigure}   
    \begin{subfigure}[b]{0.45\textwidth}
        \centering 
        \caption{Density mean relative error in space for  $u_{e_{2}}=1.065$}
        \includegraphics[width=\textwidth]{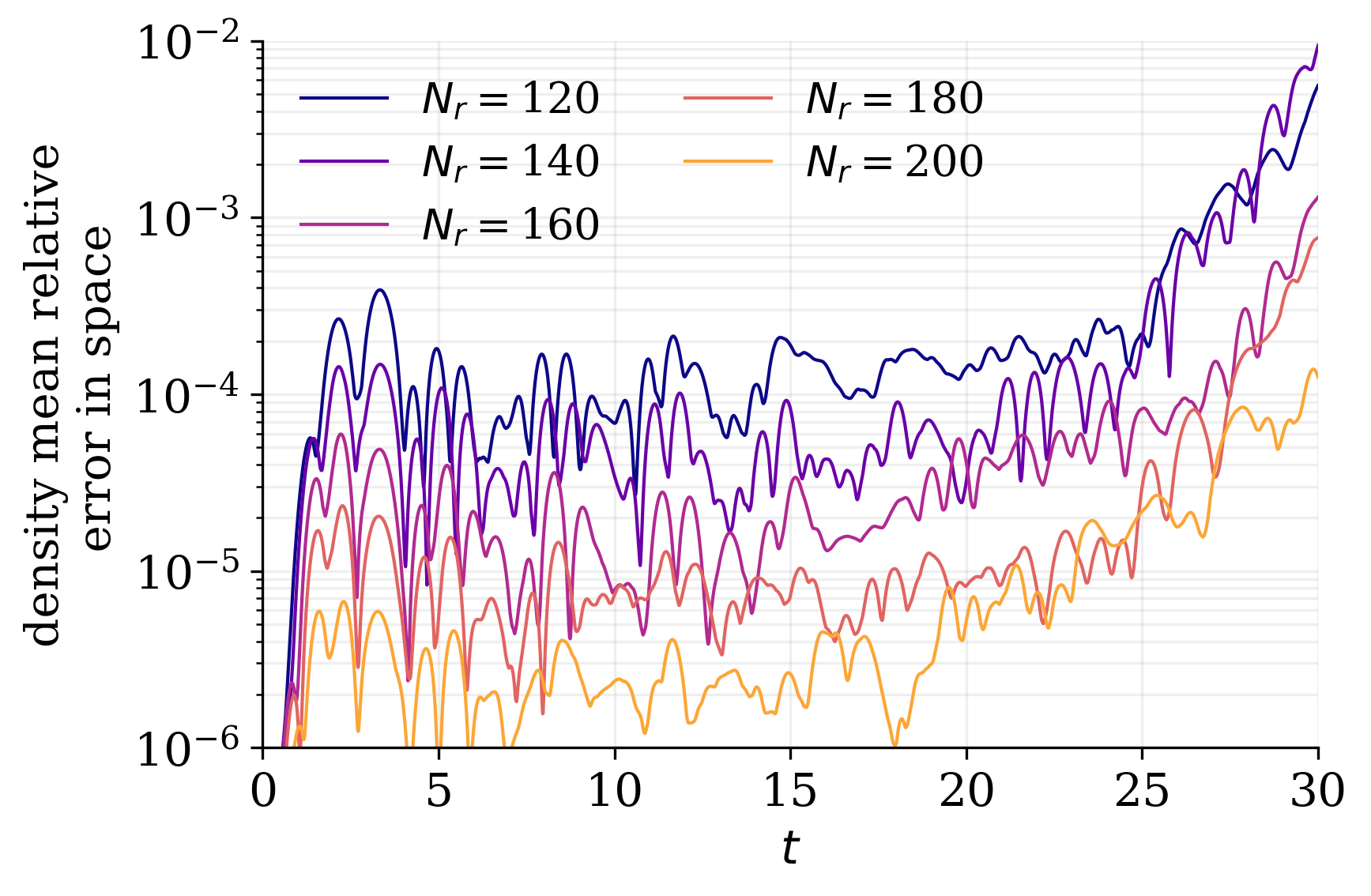}
    \end{subfigure} 
    \begin{subfigure}[b]{0.45\textwidth}
        \centering 
        \caption{Density mean relative error in space for  $u_{e_{2}}=1.09$}
        \includegraphics[width=\textwidth]{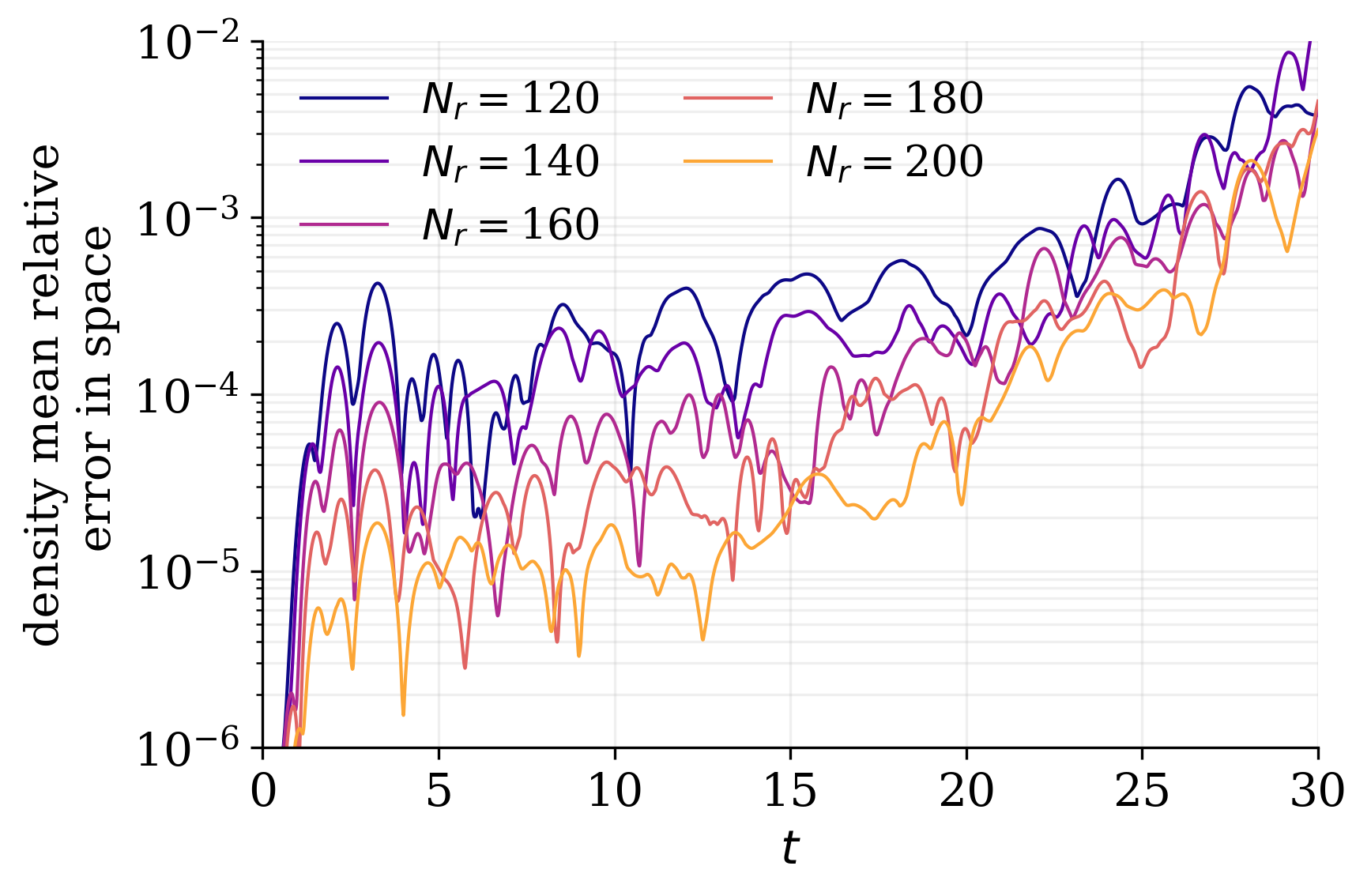}
    \end{subfigure} 
    \caption{Same as Figure~\ref{fig:weak_landau_performance} for the two-stream instability with $u_{e_{2}}=1.065$ (interpolation case) and $u_{e_{2}} = 1.09$ (extrapolation case). For $N_{r}=150$, the ROM can reduce the FOM simulation time by about a factor of $4$ with a density mean relative error in space and time of below $0.5$\% for both parametric tests. For both parametric cases, the density relative error in space grows in time as it enters the nonlinear stage of the dynamics, yet remains below 1\%. }
    \label{fig:two_stream_performance}
\end{figure}

\begin{figure}
    \centering
        \begin{subfigure}[b]{0.5\textwidth}
        \centering 
        \caption{$u_{e_{2}}= 1.09$}
        \includegraphics[width=\textwidth]{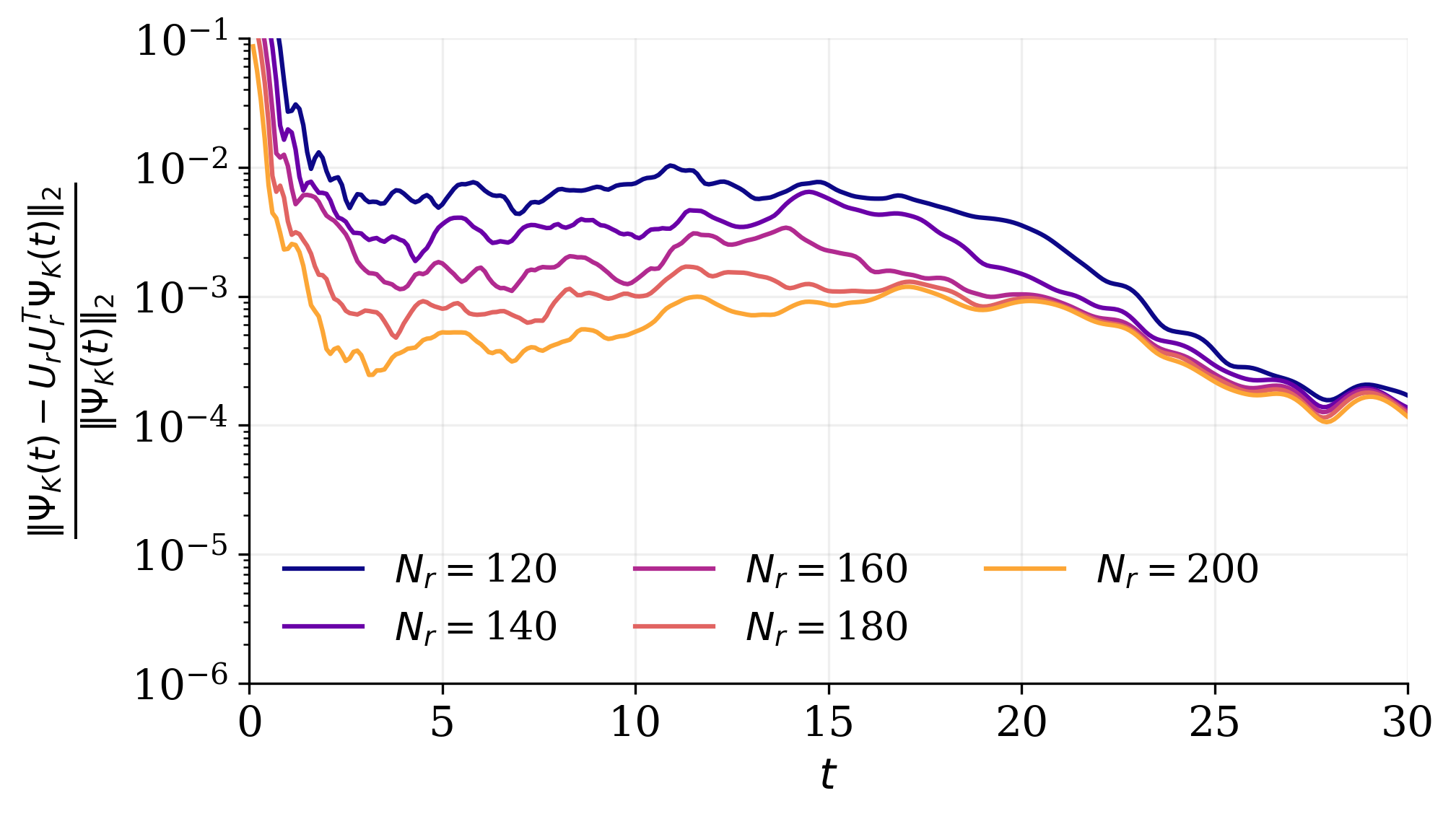}
    \end{subfigure}
    \hspace{-10pt}
    \begin{subfigure}[b]{0.5\textwidth}
        \centering 
        \caption{$u_{e_{2}} = 1.065$}
        \includegraphics[width=\textwidth]{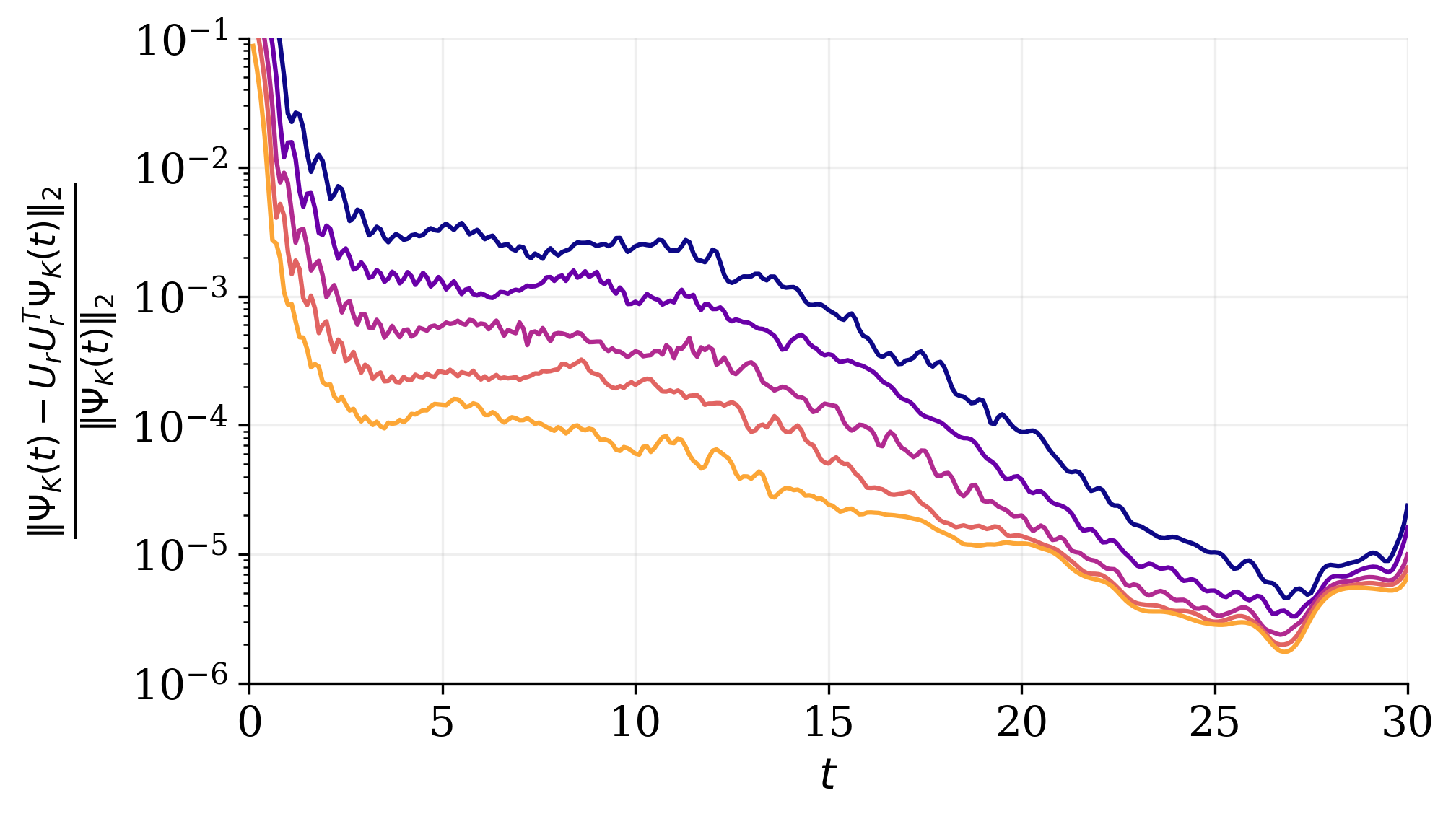}
    \end{subfigure}  
    \caption{Two-stream instability projection error with $u_{e_{2}}=1.065$ (interpolation case) and $u_{e_{2}} = 1.09$ (extrapolation case). The projection error is larger for $u_{e_{2}} = 1.09$, especially in the nonlinear saturation stage of the simulation $t> 15$. }
    \label{fig:two_stream_projection_error}
\end{figure}

We analyze the ROM's performance as the number of reduced dimensions $N_{r}$ is varied in Figure~\ref{fig:two_stream_performance}. As expected, increasing $N_{r}$ increases the CPU runtime of the ROM. The total electron density $n_{e}(x, t) = \alpha_{e_{1}} C_{e_{1}, 0}(x, t) + \alpha_{e_{2}} C_{e_{1}, 0}(x, t)$  mean relative error in space grows in time, which is unlike the electric field first Fourier mode magnitude relative error which remains fairly flat. We suspect the growth in the density mean relative error (in space) is due to the strong nonlinear behavior after $t>15$. 
Similarly, the total electron density $n_{e}(x, t) = \alpha_{e_{1}} C_{e_{1}, 0}(x, t) + \alpha_{e_{2}} C_{e_{1}, 0}(x, t)$  mean relative error in space $x \in [0, 2\pi]$ and time $t \in [0, 30]$, see Eq.~\eqref{density_mean_relative_error}, mainly decreases as a function of $N_{r}$. 
The error does not decrease monotonically since POD does not have convergence guarantees for nonlinear problems~\cite{rowley_2005_pod}. 
For the extrapolation test with $u_{e_{2}} = 1.09$, the error does not decrease exponentially as for the interpolation test with $u_{e_{1}} = 1.065$. More specifically, the error plateaus after $N_{r}=150$. This indicates that the basis is not rich enough to represent the dynamics of the extrapolation case with $u_{e_{2}} = 1.09$, see Figure~\ref{fig:two_stream_projection_error}. 
Nevertheless, for $N_{r}=150$, the ROM achieves approximately a factor $5$ speedup over the FOM simulation time while maintaining a mean relative density error in space and time below 0.5\% for both parametric tests.

\section{Conclusions}\label{sec:conclusions}
We have derived a conservative data-driven parametric ROM for the kinetic equations treated with an expansion in Hermite functions in velocity space, where the first three Hermite moments are solved exactly and the higher-order Hermite moments are approximated via POD. Due to the affine parametric dependence of the FOM, the reduced operators are constructed once \textit{a priori}, which significantly reduces the method's online computational cost. 
We examined the proposed ROM performance on 1D1V electrostatic problems: weak Landau damping and two-stream instability. 
In the weak Landau damping case, we show that the ROM successfully extrapolates both temporally and across parameters. For the more challenging two-stream instability test, which features strongly nonlinear dynamics, the ROM with $N_r = 150$ reduces the FOM CPU simulation time by approximately a factor of $5$ while maintaining a density mean relative error below 0.5\% for both parametric tests and achieving a memory reduction factor of $97$. This is an important result as it highlights the potential of complexity reduction techniques even for problems where complex structures form in the distribution function.
We anticipate that the developed ROM will be advantageous for 3D3V simulations due to its computational and memory scaling, which may enable performing multi-query parametric simulations, perhaps for design or uncertainty quantification, on laptop computers or similar devices instead of requiring supercomputer resources. 
There are several other aspects worth investigating to further improve the performance of the ROM. These include hyperreduction techniques to improve the computational scaling of the nonlinear term~\cite{chaturantabut_2010_deim, pagliantini_2025_adaptive_hyper} and the use of local bases in time to reduce $N_{r}$ for long-time nonlinear simulations~\cite{san_2015_pid, tsai_2023_rom_vp}. 
Furthermore, the number of preserved fluid moments in the ROM framework can be increased; however, for the problems analyzed in this manuscript, we found no advantage in preserving more than three moments.
The ROM can be constructed using the Legendre basis instead of the AW Hermite discretization in velocity, as both spectral expansions exhibit the fluid-kinetic coupling property~\cite{manzini_2016_legendre}. Lastly, in high-dimensional parameter spaces, full-grid sampling scales exponentially with the number of parameters, while adaptive sampling via greedy search or local sensitivity analysis can help reduce the offline computational cost~\cite{benner_2015_parametric_survey}.

\appendix
\section{Asymmetrically Weighted Hermite Basis Functions and Their Properties} \label{sec:Appendix-A}
The AW Hermite basis is defined as 
\begin{align}
    H_{n}(v; \alpha_{s}, u_{s}) &\coloneqq (\pi 2^n n!)^{-\frac{1}{2}} \mathcal{H}_{n}\left(\frac{v - u_{s}}{\alpha_{s}}\right) \exp{\left(-\frac{(v - u_{s})^2}{\alpha_{s}^{2}}\right)},  \label{hermite-basis-function-definition}\\
    H^{n}(v; \alpha_{s}, u_{s}) &\coloneqq (2^n n!)^{-\frac{1}{2}} \mathcal{H}_{n}\left(\frac{v - u_{s}}{\alpha_{s}}\right),\label{hermite-counter-basis-function-definition}\\
    \mathcal{H}_{n}(z) &\coloneqq (-1)^n \exp \left(z^2 \right) \frac{\mathrm{d}^{n}}{\mathrm{d}z^n} \exp\left(-z^2\right), \label{hermite-polynomial-definition}
\end{align}
where $\mathcal{H}_{n}$ is the \textit{physicist} Hermite polynomial~\cite{abramowitz_1964_math}.
The AW Hermite basis functions satisfy the following orthogonality relation
\begin{align}\label{orthogonal-property}
    \int_{\mathbb{R}} H_{n}(v; \alpha_{s}, u_{s}) H^{m}(v; \alpha_{s}, u_{s}) \mathrm{d} v = \alpha_{s} \delta_{n,m},
\end{align}
where $\delta_{n,m}$ is the Kronecker delta function. Additional properties of the AW Hermite basis functions that we leverage to derive Eq.~\eqref{f-vlasov-pde} are
\begin{align}
    \frac{\mathrm{d} H_{n}(v; \alpha_{s}, u_{s})}{\mathrm{d} v} &=  - \frac{2\sigma_{n+1}}{\alpha_{s}} H_{n+1}(v; \alpha_{s}, u_{s}), \label{identity-2}\\
    v H_{n}(v; \alpha_{s}, u_{s}) &= \alpha_{s} \sigma_{n+1} H_{n+1}(v; \alpha_{s}, u_{s}) + \alpha_{s} \sigma_{n} H_{n-1}(v; \alpha_{s}, u_{s})+ u_{s} H_{n}(v; \alpha_{s}, u_{s}). \label{identity-1}
\end{align}
%

\section*{Code and Data Availability}
The public repository~\url{https://github.com/opaliss/HermiteFD-ROM.git} contains a collection of Jupyter notebooks and modules in Python~3.9 with the code and data used in this study.

\section*{Acknowledgment}
O.I. was partially supported by the Los Alamos National Laboratory (LANL) Student Fellowship sponsored by the Center for Space and Earth Science (CSES). CSES is funded by LANL's Laboratory Directed Research and Development (LDRD) program under project number 20210528CR. 
O.I. was partially supported by the Strategic Enhancement of Excellence through Diversity Fellowship at the University of California, San Diego in the Department of Mechanical and Aerospace Engineering.
The LANL LDRD Program supported O.I., G.L.D., and  O.K. under project number 20220104DR. LANL is operated by Triad National Security, LLC, for the National Nuclear Security Administration of the U.S. Department of Energy (Contract No. 89233218CNA000001). 
B.K. was funded by the Applied and Computational Analysis Program of the Office of Naval Research under award N000142212624.
F.D.H. was supported by the U.S. Department of Energy, Office of Science, Office of Fusion Energy Sciences, Theory Program, under Award DE-FG02-95ER54309.

\bibliography{references}
\end{document}